\documentclass[11pt,english]{article}
\usepackage{graphicx}
\usepackage{xcolor}
\usepackage{stmaryrd}
\usepackage{amstext}
\usepackage{caption}
\usepackage{amsfonts}
\usepackage{amsmath,amssymb,epsfig}
\usepackage{amscd}
\usepackage{mathrsfs}
\usepackage[applemac]{inputenc}
\usepackage[english]{babel}
\usepackage{enumitem} 
\usepackage{subfigure}
\usepackage{hyperref}
\usepackage{blindtext}
\usepackage{verbatim}
\usepackage{cancel}
\usepackage{cases}
\usepackage{cite}
\usepackage{tensor}
\newcommand{\dd}{\mathop{}\mathopen{}\mathrm{d}}

\hypersetup{
	pdftitle={},%
	pdfauthor={},%
	pdfsubject={},%
	pdfkeywords={},%
	colorlinks=true,%
	linkcolor=blue,%
	citecolor=crimson,%
	linktocpage=true,%
	pageanchor=true}

\definecolor{crimson}{rgb}{0.7, 0.08, 0.24}

\newcommand*{\affaddr}[1]{#1}
\newcommand*{\affmark}[1][*]{\textsuperscript{#1}}

\newcommand{\be}{\begin{equation}}
\newcommand{\ee}{\end{equation}}
\def\beqa{\begin{eqnarray}}
\def\eeqa{\end{eqnarray}}
\def\bean{\begin{eqnarray*}}
\def\eean{\end{eqnarray*}}

\renewenvironment{thebibliography}[1]
         {\section*{References}\frenchspacing\small
          \begin{list}{[\arabic{enumi}]}
         {\usecounter{enumi}\parsep=2pt\topsep 0pt
         \settowidth{\labelwidth}{[#1]}
         \leftmargin=\labelwidth\advance\leftmargin\labelsep
         \rightmargin=0pt\itemsep=1pt\sloppy}}{\end{list}}

 \numberwithin{equation}{section}

\usepackage[normalem]{ulem}

\usepackage{titlesec}
\titleformat{\section}
  {\normalfont\sffamily\Large\bfseries}
  {\thesection}{1em}{}
  \titleformat{\subsection}
  {\normalfont\sffamily\large\bfseries}
  {\thesubsection}{1em}{}
  \titleformat{\subsubsection}
  {\normalfont\sffamily\large\bfseries}
  {\thesubsubsection}{1em}{}

\newcommand{\mat} [4] {\left ( \begin{array}{cc}{#1}&{#2}\\{#3}&{#4} \end{array} \right ) }

\newcommand{\p}{\partial}
\newcommand{\na}{\nabla}

\newcommand{\R}{\mathbb{R}}

\newcommand{\sgn}{\mathrm{sgn}}

\newcommand{\f}{\frac}

\renewcommand{\a}{\alpha}  \newcommand{\g}{\gamma}  
\renewcommand{\d}{\delta}  \newcommand{\epsy}{\delta P} 
 \renewcommand{\th}{\theta}   
    \renewcommand{\l}{\lambda}
\let\m=\mu    \let\n=\nu   \let\r=\rho 
 \newcommand{\s}{\sigma}  \renewcommand{\t}{\tau}    
     
 \let\Om=\Omega
\newcommand{\sscr}{\scriptscriptstyle\rm}
\newcommand{\Lf}{\ell_0}
\newcommand{\os}[1]{\overset{\circ}{#1}}

\title{\textbf{\textsf{Generic features of a polymer quantum black hole}}\vspace{0.25cm}}

\author{
\textsf{Johannes M\"unch\affmark[1]\footnote{\texttt{johannes.muench@cpt.univ-mrs.fr}}$\,$,$\,$ Alejandro Perez\affmark[1]\footnote{\texttt{alejandro.perez@cpt.univ-mrs.fr}} $\,$,$\,$ 
Simone Speziale\affmark[1]\footnote{\texttt{simone.speziale@cpt.univ-mrs.fr}} $\,$,$\,$ 
Sami Viollet\affmark[1]\footnote{\texttt{sami.viollet@cpt.univ-mrs.fr (corresponding author)}}}\\
\affaddr{\affmark[1]\textsf{Aix-Marseille Universit\'e, Universit\'e de Toulon, CNRS, CPT,}}\\
\affaddr{\textsf{13288 Marseille, France}}\vspace{-0.5cm}
}

\begin{document}
%--------------------------------------------------------------------
%--------------------------------------------------------------------

\begin{sffamily}
	\maketitle
\begin{abstract}
	\noindent
	\textsf{Non-singular black holes models can be described by modified classical equations motivated by loop quantum gravity. We investigate what happens when the sine function typically used in the modification is replaced by an arbitrary bounded  function, a generalization meant to study the effect of ambiguities such as the choice of representation of the holonomy. A number of features can be determined without committing to a specific choice of functions. We find generic singularity resolution. The presence and number of horizons is determined by global features of the function regularizing the angular components of the connection, and the presence and number of bounces by global features of the  function regularizing the time component. The trapping or anti-trapping nature of regions inside horizons depends on the relative location with respect to eventual bounces. We use these results to comment on some of the ambiguities of polymer black hole models.}
\end{abstract}	

\tableofcontents
\vspace{-0.25cm}
\end{sffamily}

%--------------------------------------------------------------------
\section{Introduction}
%--------------------------------------------------------------------

An emblematic problem in quantum gravity is to understand the fate of the black hole singularity predicted by general relativity. 
Aside from extremely simplified 2d models, see e.g. \cite{Callan:1992rs,Ashtekar:2010hx}, explicit calculations are not within current reach, and this motivates the investigation of minisuperspace models in order to shed useful light on qualitative aspects of the process. 
In loop quantum gravity (LQG), quantum minisuperspace models are constructed using the key input from the full theory that the fundamental quantum operators are holonomies of the Ashtekar-Barbero connection. These are defined in a representation which lacks weak-continuity, thus making the models unitary inequivalent to those based on metric variables. A compelling result of this approach is the generic singularity-avoidance \cite{Ashtekar:2006wn}, with the minisuperspace dynamics predicting a bounce occurring at a critical energy or curvature density. This is valid in both cosmological \cite{AshtekarLoopquantumcosmology:astatusreport,Ashtekar:2006wn} and black hole \cite{Kelly:2020uwj,AchourNonSingularBlackHoles,AshtekarQuantumTransfigurarationof,BodendorferEffectivequantumextended,Alesci:2020zfi} models. In the latter case, this type of investigations first started in \cite{Modesto:2005zm,Ashtekar:2005qt}.

One key aspect of the construction is that the resulting quantum corrections can be equally predicted using an effective classical dynamics, where the original Hamiltonian is modified in a precise way \cite{TaverasCorrectionstothe}. This step is referred to as polymerization, and allows one a simple description of the system, and an independent exploration of this type of models.
In spite of the compelling singularity resolutions, these models are only a first step towards a clear understanding of the non-perturbative quantum gravitational effects. To move forward, it is important to address their limitations and shortcomings. For instance, the models rely on a specific foliation, and restoring foliation-independence is far from accomplished \cite{Bojowald:2022zog, Bojowald:2021isp, Bojowald:2020unm, Bojowald:2019dry}. 
The choice of polymerization scheme is not identified a priori from fundamental principles or deductions from the full theory; but rather reconstructed a posteriori requiring a good semiclassical limit at large scales, and there is no uniqueness about the procedure used \cite{Amadei:2022zwp}.
We will not have much to say about the first issue, but we would like to focus on the following aspect about the second. The function used in the scheme is typically picked to be a sine. This is supposed to represent working with the holonomy in the fundamental representation in full loop quantum gravity. However it has been known for a while that there may be quantization ambiguities associated with this choice already in the full theory \cite{Perez:2005fn}.\footnote{It has been recently pointed out in \cite{Varadarajan:2021zrk} that some of the ambiguities can be reduced by a new quantization of the Hamiltonian constraint of the full theory; nonetheless, ambiguities of the type analysed here remain in the part of the Hamiltonian constraint responsible for non-trivial propagation \cite{ale-madhavan}.} These ambiguities affect models like \cite{Alesci:2019pbs} in which the polymerization scheme is derived from the full theory taking expectation values with respect to symmetry-reduced coherent states. The effect of such ambiguities where recently analysed in the context of quantum cosmology in \cite{Amadei:2022zwp}, showing that they strongly affect the physics; here, we investigate how changes in the choice of polymer function affect static and spherically symmetric black hole models. 

To that end, we focus on a specific polymer black hole model, the one proposed in \cite{BodendorferEffectivequantumextended} that we refer to as the BMM (Bodendorfer, Mele, Münch) model from here on. 
The model is a spherically symmetric black hole, with a four-dimensional phase space. The two configuration variables correspond to a time and an angular component of the metric. The model has the limitations mentioned above: 
It is based on a fixed foliation, and uses a specific choice of variables to be parametrized, identified because they realise a so-called $\bar\m$-scheme on the Ashtekar-Barbero variables. We will not touch these choices. We will instead investigate what happens if the sine functions used in the polymerization \cite{BodendorferEffectivequantumextended} are replaced by \emph{arbitrary} functions. 
We expect that our techniques and the implications of our results can be applied to models of polymer quantum black holes such as \cite{VakiliClassicalpolymerizationof,CorichiLoopquantizationof,ModestoSemiclassicalLoopQuantum,BoehmerLoopquantumdynamics,BenAchourPolymerSchwarzschildBlack,AshtekarQuantumExtensionof,AshtekarQuantumTransfigurarationof,BodendorferEffectivequantumextended,BodendorferAOSNote,BodendorferMassandHorizon,Bodendorferbvtypevariables,AssanioussiPerspectivesonthe,KellyEffectiveloopquantumgravity,KellyBlackholecollapse,GambiniSphericallysymmetricloop,GeillerSymmetriesofthe,SartiniQuantumdynamicsof,Bouhmadi-LopezAsymptoticnon-flatness,GanPropertiesofthe,Garcia-QuismondoExpolringalternativestothe,GambiniLQBHextensions,ElizagaNavascues:2022rof} other than BMM.

At first sight, it may seem better to just consider explicit alternatives, e.g. changing and/or superimposing frequencies and phase shifts, in order to mimic the use of different irreducible representations of the holonomies and of different regularizing paths. But it turns out that many properties of the polymer black holes are accessible without making an explicit choice for these functions.
%it is possible to compute all relevant features of the polymer black hole for \emph{arbitrary} polymer functions. 
This remarkable fact is due to the simplicity of the model, in particular the fact that one of the two Dirac observables remains simple even after polymerization. The second does not, and it makes some of ours formulas implicit, without however hindering our considerations. 

The two polymer functions that we keep arbitrary correspond to respectively the time and angular component of the connection. We show first of all that the configuration variables of \cite{BodendorferEffectivequantumextended} produce a $\bar\m$-scheme for any choice of polymer functions. Requiring the correct semiclassical limit at spatial infinity imposes a condition on the first derivatives of both polymer functions, but also a condition on the second derivative of the angular polymer function. Regularity of both functions avoids singularities, and replaces them with bounces. The number of bounces turns out to be determined by the angular function alone; whereas the number of horizons is determined by the time function alone. Their relative location depends on both the chosen functions and the solution considered. We provide a general graphical analysis to deduce these properties without committing to specific choices, but our analysis and formulas can be of help also in studying a specific model that one may be interested in.

Our results shows that there is a valuable richness in the class of polymer black holes, and that considerable mathematical control can be kept also relaxing the standard choice of sine polymer functions. We hope that some of this control can be used to address some of the limitations and help constructing more robust models.

We use units $G=c=1$.

%--------------------------------------------------------------------
\section{From the classical black hole to a polymerized black hole }
%--------------------------------------------------------------------
We will review the classical Schwarzschild solution, based on the Hamiltonian formulation, in order to introduce the notations commonly used in the literature. More precisely, after giving the general expression of the line element compatible with the symmetry considered (i.e. spherical symmetry and staticity) in term of the geometrodynamic variables. Then we will recall how the problem can be reformulated in terms of Ashtekar-Barbero variables and how this reformulation can lead to a new convenient set of variables, the $(v,P)$ variables. From there, we will quickly recall the general ideas of the BMM polymerisation model which will be useful in the next part. We will highlight the main result of this model like the resolution of the singularity but we will also insist on the limitation of this model caused by the particular choices on which it is built.

%--------------------------------------------------------------------
\subsection{Minisuperspace black hole model} 
%--------------------------------------------------------------------

We consider the following spherically symmetric and static ansatz
\begin{equation}\label{line1}
ds^2=-\bar a(r)dt^2+\f{\bar n(r)}{\bar a(r)}dr^2+\bar b^2(r)d\Om^2.
\end{equation}
The spatial diffeo constraint vanishes identically, and we eliminated the shift vector via our choice for the coordinates $t, \th$ and $\phi$. On the other hand, we have not fixed the $r$ coordinate to be the area radius, and there is still a non-trivial Hamiltonian constraint to be solved associated with this reparametrization.
Plugging this ansatz into the Einstein's equations, one finds
\be\label{Lsol}
\bar n(r) = c_1^2 \dot {\bar b}(r)^2, \qquad \bar a(r) =c_1^2\left(1-\frac{c_2}{b(r)}\right).
\ee
The solution space is thus parametrized by two constants of integration, and $\bar n(r)$ is arbitrary. However only one of the constants has a geometrical interpretation, since $c_1$ can always be reabsorbed by rescaling the coordinate $t$. The metric in this form can be recognized to be the Schwarzschild solution, with 
\be\label{Mc2}
M=\f{c_2}2,
\ee
asymptotic unit proper time $\t:=c_1 t$ and area radius $b$. We can freely choose a constant $\bar n(r)\equiv c_1^2$, and this will make $r$ the area radius.  In other words, one constant of integration is the black hole's mass, another is a rescaling of the asymptotic time, and $\bar n$ is the freedom of $r$-reparametrizations.

The symmetry reduction makes the Einstein-Hilbert action, as well as the ensuing Poisson structure, divergent on a open topology. In the following, we will study the canonical structure of the foliation by hypersurfaces of constant $r$. This is because $r$ is a time-like coordinate inside the horizon, and we are ultimately interested in the dynamics near the singularity. 
To regularize the action, one introduces a fiducial cell for the coordinates $x:=(t,\th,\phi)$ given by $C:=[0,\Lf]\times S^2$. 
The physical size $L_0$ of the fiducial cell is $r$-dependent. Taking for instance spatial infinity as reference, we have
\be\label{Cphys}
L_0:=\int_0^{\Lf} \sqrt{|g_{tt}(r=\infty)|}dt = \Lf |c_1|
\ee 
on-shell.
Then, 
\begin{equation}\label{action}
S[C]=\frac{1}{16\pi}\int dr\int_{C} d^3x\sqrt{-g}R +{\rm boundary \ term}
= \int dr \, L, 
\end{equation}
where 
\be\label{defL}
L:= \frac{\Lf\sqrt{\bar n}}{2}\left(\frac{\dot {\bar a} \dot{ \bar b} \bar b}{\bar n}+\frac{\bar a \dot{\bar b}^2}{\bar n}+1\right).
\ee
 The boundary term to be added to the action is the Gibbons-Hawking-York term, which takes care of eliminating derivatives on lapse and shift.
For the interested reader, details are reported in Appendix \ref{Boundary term}.
It is convenient to reabsorb the coordinate length $\Lf$ changing  variables to:
\begin{align}\label{varredef}
\sqrt a:= \Lf \sqrt{\bar a}, \qquad 
b:=\bar b, \qquad
\sqrt{n}:=\Lf \sqrt{\bar n},
\end{align}
Notice that $\sqrt{a}$ is the physical length of the fiducial cell in the $t$ direction as a function of $r$. We end up with the following Lagrangian,
\begin{equation}
L( a, b, n)=\frac{\sqrt{ n}}{2}\left(\frac{\dot {a} \dot{ b}  b}{n}+\frac{ a \dot{ b}^2}{ n}+1\right)
\end{equation}
which will give a well-defined phase space.

At this point the spherical black hole minisuperspace model is effectively being described by a point-particle mechanical system.\footnote{And it is part of a class of minisuperspace models described by the motion of point particles in a given supermetric, which in this case reads
\[
G=\f{2\sqrt n}b\mat{-a/b}{1}{1}{0}. 
\]
See e.g. \cite{Geiller:2022baq} for a description in these terms of generic spherical black holes and Bianchi models.}

The solution space is the same as before, in particular
\be
a=\Lf^2c_1^2\left(1-\f{c_2}b\right).
\ee 
However, there is an important difference about the physical interpretation of the solutions:
in the point-particle description we have no longer access to the coordinate $t$, hence it is not possible to reabsorb $c_1$ with a time diffeomorphism.
In other words, fiducial cells of different physical size \eqref{Cphys} correspond to physically distinct points in the solution space.

Turning $c_1$ from an irrelevant constant to a physical one is clearly an artefact of the regularization, but it is also natural in the following sense. In the full space space, the Schwarzschild solution is a one-parameter trajectory. The smallest phase space in which this family can be embedded is 2-dimensional, hence we expect a second variable to be present and to be conjugated to the mass, and $c_1$ plays this role.
The role of the regularization in defining this two-dimensional phase space can also be nicely understood looking at the symplectic potential obtained from the on-shell variation of \eqref{action}, which in the area-radius gauge gives
\be
\theta=-\frac{1}{2} \ell_0
   \left(c_1 {dc_2}+2
   {c_2} {dc_1}-2 r {dc_1}\right),
\ee
with symplectic structure
\be\label{esta}
\omega=\frac{\ell_0}{2} 
   dc_1 \wedge {dc_2}.
\ee
Removing the regulator taking $\ell_0\to \infty$ gives a divergent symplectic structure. One can also see that the commutation relation $\{c_1,c_2\}=2/\ell_0$ has a vanishing limit, and interpret this as a sign that one of the coordinates loses its dynamical role, $c_1$ in this case, and can be reabsorbed into the redefinition of asymptotic time .\footnote{This is the same observation already made at the quantum level in \cite{Rovelli:2013zaa}.}

We now review the Hamiltonian of the system associated with the foliation by $r=$constant hypersurfaces. The signature of this foliation is not fixed, being time-like outside the horizon, null at the horizon, and space-like inside. No issues arise from this fact, thanks to the symmetry reduction the dynamics is well-defined for all values of $r$. The lapse function in the interior is
\be\label{lapse}
N(r) = \f{\sqrt{n(r)}}{\sqrt{-a(r)}}.
\ee
Inspection of the Lagrangian shows that one can freely trade between $N$ and $n$ as Lagrange multipliers.
With this choice of time, the conjugate momenta with respect to this global time parameter are
\begin{align}
p_a=\frac{\partial L}{\partial \dot a}=\frac{b\dot b}{2\sqrt{n}}, \qquad 
p_b=\frac{\partial L}{\partial \dot b}=\frac{2a\dot b +\dot a b}{2\sqrt{n}}, \qquad 
p_n=\frac{\partial L}{\partial \dot n}=0.
\end{align}
The last equation gives a primary constraint, that should be added to the Legendre transform of the Lagrangian. 
Upon doing so, one finds the primary Hamiltonian
\begin{equation}
H(n)=\sqrt{n}\left(\frac{2p_ap_b}{b}-\frac{2ap_a^2}{b^2}-\frac{1}{2}\right)+\lambda p_n,
\end{equation}
where $\lambda$ is a Lagrange multiplier. 
Stabilizing the primary constraint, i.e. imposing $\dot p_n \approx 0$, leads to the secondary constraint 
\begin{equation}
\psi:=\frac{2p_ap_b}{b}-\frac{2ap_a^2}{b^2}-\frac{1}{2}\approx 0.
\end{equation}
This is the Hamiltonian constraint generating $r$-diffeomorphisms. It is automatically stable, and no further constraints arise in the analysis. We thus obtain a 2-dimensional physical phase space, in agreement with the two-parameter family of solutions \eqref{Lsol}. 
Moreover $\dot n = \lambda$, hence %just like in the full theory 
we can treat $\sqrt{n}$ as a Lagrange multiplier, remove the pair $(n,p_n)$ from the phase space and set $\l=0$ without any loss of generality. Then we have a 4d kinematical phase space with Hamiltonian constraint $H(n)=\sqrt{n}\psi$.
We will use this formulation from now on.

Consider now the following canonical transformation \cite{BodendorferEffectivequantumextended}, 
\begin{subequations}\label{cantransf}
\begin{align}
& v_1:=\frac{2}{3}b^3, \qquad  \ P_1:=
\frac{p_b}{2b^2}-\frac{ap_a}{b^3}=\frac{\dot a}{4\sqrt{n}b}, \label{v1(b)}\\
& v_2:=2ab^2, \qquad  P_2:=
\frac{p_a}{2b^2}=\frac{\dot b}{4\sqrt{n}b}, \label{v2(a)}\\
& \{v_i, P_j\}=\d_{ij}.\label{vPpb}
\end{align}
\end{subequations}
It is the generalization to the black hole model of a similar transformation used in loop quantum cosmology, and it will be relevant to construct the effective quantum model below.
In terms of the new variables the Hamiltonian constraints becomes
\begin{equation} \label{HvP}
H(n)=\sqrt{n}\left(12 v_1 P_1 P_2+ 4v_2 P_2^2-\frac{1}{2}\right),
\end{equation}
and generates the dynamical equations
\begin{subequations}
\begin{align}\label{classical eom}
\dot v_1&=12\sqrt{n}v_1P_2, \\
\dot v_2&=12\sqrt{n}v_1P_1+8\sqrt{n}v_2P_2,\\
\dot P_1&=-12\sqrt{n}P_1P_2,
\label{eq c}\\
\dot P_2&=-4\sqrt{n}P_2^2.
\label{eq d}
\end{align}
\end{subequations}
Notice that the equations for $P_i$ are decoupled. It is also possible to easily identify two constants of motion. The first can be deduced by inspection of the Hamiltonian, and it is given by 
\begin{equation}\label{K1}
K_1=v_1P_1.
\end{equation}
The second can be found taking the ratio of \eqref{eq c} and \eqref{eq d} and integrating, giving
\be\label{K2}
K_2=\ln{|P_1|}-3\ln{|P_2|}.
\ee
In performing the integration we fix the integration constants so that the two contributions to $K_2$  vanish when $P_1=1$ and $P_2=1$. 
The two constants of motion turn out to be canonically conjugated functions, $\{K_1,K_2\}=1$, and commute with the Hamiltonian constraint by construction. Therefore they provide two Dirac observables for the system. 

It is instructive to derive the Schwarzschild solution from this formulation of the dynamics, because it will help comparison with the polymerized model below.
To solve the equations, let us fix its gauge freedom imposing 
$n=n_o$ constant, Then, \eqref{eq d} is solved by
\begin{equation}\label{P2 classical}
P_2(r)=\f1{4\sqrt{n_o}r}, 
\end{equation}
up to a constant of integration that can always be absorbed by a shift of the $r$ coordinate. This solution is valid provided $r\neq 0$. We restrict the domain to $r>0$, hence $b>0$. Then, using \eqref{K2} directly leads to
\begin{equation}
P_1(r)=\textrm{sign}(P_1)\frac{e^{K_2}}{(4\sqrt{n_0}r)^3}.
\end{equation}
From this, \eqref{K1}, and the fact that $v_1=\frac{2}{3}b^3>0$, we derive that
\begin{equation}\label{v1 classical}
P_1(r)=\frac{\textrm{sign}(K_1)e^{K_2}}{(4\sqrt{n_0}r)^3} \quad \textrm{and} \quad v_1(r)=(4\sqrt{n_o})^3 |K_1| e^{-K_2}r^3,
\end{equation}
Finally, from the vanishing of the Hamiltonian constraint \eqref{HvP} we derive that
\begin{equation}
v_2(r)=4\sqrt{n_o} \left(\frac{1}{2}\sqrt{n_o}-\frac{3K_1}{r}\right)r^2.
%\qquad v_2(r)=\mathscr{L}_or^2\left(\frac{1}{8}\mathscr{L}_o-\frac{3CD}{r}\right).
\end{equation}

This general solution is of course equivalent to \eqref{Lsol} found earlier. To see this explicitly, we can reconstruct the metric. Let us do so in terms of the unbarred quantities \eqref{varredef}, in terms of which it reads
\begin{equation}\label{eq:canonicalmetric2}
	\dd s^2 = -\frac{a(r)}{\Lf^2} \dd t^2 + \frac{n(r)}{a(r)} \dd r^2 + b(r)^2 \dd \Om^2.
\end{equation}
Inverting \eqref{cantransf}, we obtain
\begin{align}
a&= \frac{e^{2K_2/3}}{16}\left(\frac{2}{3|K_1|}\right)^{2/3} \left(1- \frac{24 \left(\frac{3}{2}\right)^{1/3}K_1|K_1|^{1/3}e^{-K_2/3}}{b}\right),
\\
b&=4\sqrt{n_o}\left(\frac{3|K_1|}{2}\right)^{1/3}e^{-K_2/3} r .
\label{b classical}
\end{align}
where
\be\label{MCD}
M=2^{5/3}3^{4/3}K_1|K_1|^{1/3}e^{-K_2/3}.   
\ee

We see that the metric describes the Schwarzschild solution with mass $M$, area radius $b$, and asymptotic time 
\be \label{asymptotic time}
t_S:=\frac{e^{K_2/3}}{4\Lf}\left(\frac{2}{3|K_1|}\right)^{1/3} t. 
\ee
The solution space is spanned by all values $K_{i}\in\R^2$. Each point in this space describes a Schwarzschild black hole (with given mass---positive or negative---asymptotic time and area radius), except the line $K_1=0$ which describes a degenerate spacetime. The zero-mass limit %Minkowski solution is
corresponds to the limit $K_i\rightarrow(0^\pm,-\infty)$. 

The last equation \eqref{b classical} shows that  if one wants to fix the gauge $b=r$, then the constant $n_0$ has to be chosen in a phase-space dependent manner (note that we had already stumbled upon this fact below equation \eqref{Mc2})\footnote{There is a subtlety about $n_0$, not relevant for the following, but useful to make contact with a discussion made in \cite{BodendorferEffectivequantumextended}. Under a change of fiducial cell $\Lf \mapsto \alpha \Lf$, the function $n$ rescales as $n\mapsto \a^2 n$. Under a time diffeomorphism $t\mapsto t'(t)$, it is invariant---this follows from its definition
	\[
	\sqrt{n}:=\Lf \sqrt{\bar n} =\int_0^{\Lf}\sqrt{-g_{tt} g_{rr}} dt
	\]
	which makes it akin to a volume quantity. Therefore, also the constant $n_0$ has to be chosen in an $\Lf$-dependent way in order to preserve this properties.
	
}.
For convenience, we also report the relation between the $K_i$'s and the $c_i$'s, 
\be
K_1= \frac{\Lf c_1 c_2}{6}\ \ \ \ K_2=\log\left |16\Lf^4c_1^4c_2\right |=\log\left | 16\Lf^4c_1c_2\right |+3\log{|c_1|}.
\ee
From these relations we recover the results of the previous section, in particular the expression \eqref{Mc2} for the mass, 
and $c_1$ being the only parameter that enters the rescaling of the asymptotic time.

Let us add a few remarks useful in the following.
\begin{itemize}
\item
On the interpretation of the $P_i$ variables. Using the on-shell values 
\eqref{P2 classical} and \eqref{v1 classical} and the relation between $b$ and $r$ \eqref{b classical}, one finds \cite{BodendorferEffectivequantumextended}
\begin{align}
P_1(b)&=\frac{M}{b^3}\left(\frac{2}{3|K_1|}\right)^{1/3}\frac{e^{K_2/3}}{8},\qquad
P_2(b)=\frac{1}{b}\left(\frac{3|K_1|}{2}\right)^{1/3}e^{-K_2/3}.
\end{align}
Thus $P_1$ is proportional to the (square root of the) Kretschmann scalar, and $P_2$ to the inverse area radius of the Schwarzschild black hole.

\item Any phase space function that is monotonic in $r$ will be a good clock for our Hamiltonian evolution.
From the general solution, we see that all variables except $v_2$ are monotonic functions of $r$. 
Therefore, all of them provide good internal clocks. If we choose $P_2$ in particular, we can rewrite the metric in a lapse-independent way,
\begin{align}
ds^2=&-\f1{16}\left( \f{2e^{K_2}}{3K_1} \right)^{2/3} \left(1 - {24 K_1P_2} \right)\dd\tau^2
\notag
\\
&+\left(\frac{3K_1\lambda_1}{2\lambda_1P_1}\right)^{2/3} \left(1 - {24 K_1P_2} \right)^{-1} \l_2{}^2 \dd P_2{}^2+\left(\frac{3K_1\lambda_1}{2\lambda_1P_1}\right)^{2/3}\dd\Omega^2, \label{Karim} \\
\tau:=&\f{t}{\Lf}
\end{align}
This is the form of the metric that we will write below for the polymerized model.

\end{itemize}

%--------------------------------------------------------------------
\subsection{Ashtekar-Barbero variables}\label{AB variables}
%--------------------------------------------------------------------

The first step towards polymerisation is a canonical transformation to Ashtekar-Barbero variables,
\be
A_a^i:= \Gamma^i_a+\gamma K^i_a, \qquad E^a_i = \frac{1}{2}\epsilon_{ijk}\epsilon^{abc}e_b^je_c^k,
\ee
where $\gamma$ is the Immirzi parameter.
For the interior of a spherically symmetric and static spacetime we can take them to be\cite{AshtekarQuantumExtensionof,AshtekarQuantumTransfigurarationof,KellyEffectiveloopquantumgravity}\footnote{With the renaming $b(r)\mapsto d(r)$ to avoid confusion with our $b(r)$ defined in \eqref{line1} and \eqref{varredef}.}

\begin{align}
A_a^i\tau_idx^a&=\frac{c(r)}{\Lf}\tau_3 dt+d(r)\tau_2 d\theta-d(r) \tau _1 \sin\theta d\phi +\tau_3 \cos\theta d\phi, \\
E^a_i\tau^i\partial_a&=p_c(r)\tau_3 \sin\theta \p_t+\frac{p_d(r)}{\Lf}\tau_2 \sin \theta \partial_{\theta}-\frac{p_d(r)}{\Lf}\tau_1 \partial _{\phi},
\end{align}
where $\tau_i=(-i/2) \sigma_i $ and $\s_i$ are the Pauli matrices.
The map to the previous variables is given by
\begin{subequations}\label{AB to vP}
\begin{align}
\textrm{sign}(p_c) \, c&=-2\g \left(\frac{p_b}{2b}-\frac{ap_a}{b^2}\right)=-2\g P_1\left( \frac{3v_1}{2}\right)^{1/3}, \\
\textrm{sign}(p_d)\, d&=2\g \frac{\sqrt{-a}p_a}{b}=4\g P_2\sqrt{-\frac{v_2}{2}},\\
\textrm{sign}(p_c)\, p_c&=b^2=\left(\frac{3v_1}{2}\right) ^{2/3},\\
\textrm{sign}(p_d)\, p_d&=\sqrt{-a}b= \sqrt{-\frac{v_2}{2}}.
\end{align}
\end{subequations}
The overall sign ambiguity is the usual global parity freedom in the triad, whereas the relative sign ambiguity is a consequence of the symmetry reduction, and cannot be eliminated \cite{Ashtekar2006}.
This map is a canonical transformation, with non vanishing Poisson brackets given by
\begin{subequations}
\begin{align}
\{c,p_c\}=2\g, \qquad \{d,p_d\}&=\g,
\end{align}
\end{subequations}
and Hamiltonian constraint
\begin{equation}\label{Hcp}
H(N)=-\frac{Nd}{2\gamma^2 \, \textrm{sign}(p_c)\sqrt{|p_c|}}\left(2c p_c +\left(d+\frac{\gamma ^2}{d}\right)p_d\right).
\end{equation}
As before lapse is related to our choice of Lagrange multiplier $n$ by \eqref{lapse}, a redefinition which does not affect the equations of motion.\footnote{Hamilton's equations are modified by the addition of a term proportional to the constraint and thus zero on-shell.}

%-----------------------------------------------
\subsection{BMM polymerisation} \label{BMMp}
%-----------------------------------------------

The BMM model \cite{BodendorferEffectivequantumextended} is based on the $(v_i,P_i)$ phase space variables we reviewed above. There are two key observations made in \cite{BodendorferEffectivequantumextended}  that motivate the use of these variables. The first is that as we observed above, $P_1$ is related to the Kretschmann scalar and $P_2$ to the inverse area radius. Hence we could intuitively expect that replacing these variables by bounded functions should remove the classical divergences of the dynamics. The second is that polymerizing these variables with a non-dynamical regulator induces a $\bar\m$-scheme on the Ashtekar-Barbero variables, compatible with the Hamiltonian. Hence these variables are the analogue of the $(v,b)$ variables in loop quantum cosmology. To see this, consider the replacement
\begin{align}\label{polymerisation}
P_i&\rightarrow \frac{\sin(\lambda_iP_i)}{\lambda_i}, \qquad i=1,2,
\end{align}

where $\lambda_1$ and $\lambda_2$ are a priori independent polymerisation scales (i.e. the scale at which the modification of the dynamic will be important). 
We take them to be constants, and of Planckian order. 
From \eqref{AB to vP} we have that
\begin{align}\label{eq:schemecb}
&\lambda_1P_1=\delta_cc, \qquad \delta_c:=-\textrm{sign}(p_c)\frac{\lambda_1}{2\gamma\sqrt{|p_c|}},\\
&\l_2P_2=\delta_dd, \qquad \delta_d:=\textrm{sign}(p_d)\frac{\lambda_2}{4\gamma|p_d|}.
\end{align}
Hence the replacement \eqref{polymerisation} is equivalent to 
\begin{align}\label{mubar}
&c\rightarrow\f{\sin (\delta_c c)}{\d_c}, \qquad d\rightarrow\f{\sin(\d_d d)}{\d_d d},
\end{align}
where the quantum parameters $\delta_c$ and $\delta_d$ have a phase-space dependance given by \eqref{eq:schemecb}.
This phase-space dependence {can be used to introduce the area-gap of the full theory}, thus defining a $\bar\m$-scheme for the Ashtekar-Barbero variables. The replacement \eqref{polymerisation} leads to the modified Hamiltonian constraint 
\begin{equation}\label{eq:HeffvP}
H_{\text{eff}}(n)=\sqrt{n}\mathcal H_{\text{eff}} \quad , \quad \mathcal H_{\text{eff}} = 12v_1\frac{\sin(\lambda_1P_1)}{\lambda_1}\frac{\sin(\lambda_2 P_2)}{\lambda_2}+4v_2\frac{\sin(\lambda_2 P_2)^2}{\lambda_2^2}-\frac{1}{2}\approx 0.
\end{equation}
On the other hand, the polymerization of the Ashtekar-Barbero Hamiltonian \eqref{Hcp} under \eqref{mubar} gives
	\begin{equation*}
		H_{\text{eff}}(N)=-\frac{N \sin\left(\delta_d d\right)}{ \delta_d 2\gamma^2 \, \textrm{sign}(p_c)\sqrt{|p_c|}}\left(2\frac{\sin\left(\delta_c c\right)}{\delta_c} p_c +\left(\frac{\sin\left(\delta_d d\right)}{\delta_d}+\frac{\gamma ^2 \delta_d}{\sin\left(\delta_d d\right)}\right)p_d\right).
	\end{equation*}
Inserting
\eqref{eq:schemecb}, \eqref{AB to vP} and \eqref{lapse} in the latter recovers the former, therefore the polymerization \eqref{polymerisation} is consistent with the $\bar\m$-scheme also at the dynamical level.

The dynamics generated by the polymerized Hamiltonian is {expected to capture the mean field quantum corrections \cite{TaverasCorrectionstothe} and it is thus referred to as effective dynamics.} This effective dynamics is
obtained from \eqref{eq:HeffvP}  and the original Poisson brackets \eqref{vPpb}, giving
\begin{subequations}\label{polymerised equations}
\begin{align}
\dot v_1&=12\sqrt{n}v_1\cos(\lambda_1P_1)\frac{\sin(\lambda_2P_2)}{\lambda_2} \ ,\\
\dot v_2&=12\sqrt{n}v_1\frac{\sin(\lambda_1P_1)}{\lambda_1}\cos(\lambda_2P_2)+8\sqrt{n}v_2\frac{\sin(\lambda_2P_2)}{\lambda_2}\cos(\lambda_2P_2) \ ,\\
\dot P_1&=-12\sqrt{n}\frac{\sin(\lambda_1P_1)}{\lambda_1}\frac{\sin(\lambda_2 P_2)}{\lambda_2} \ ,\\
\dot P_2&=-4\sqrt{n}\frac{\sin(\lambda_2P_2)^2}{\lambda_2^2} \ ,\\
\mathcal H_{\text{eff}}&= 3v_1\frac{\sin(\lambda_1P_1)}{\lambda_1}\frac{\sin(\lambda_2 P_2)}{\lambda_2}+v_2\frac{\sin(\lambda_2 P_2)^2}{\lambda_2^2}-\frac{1}{8} = 0 \;.
\end{align}
\end{subequations}
It was shown in \cite{BodendorferEffectivequantumextended} that the equations can be solved analytically, and the following properties ensue:
\begin{enumerate}
\item The singularity at $b=0$ is removed.
\item There are two asymptotic spacetime regions, both approaching a Schwarzschild geometry.
\item There is a space-like surface at some $b_{min}\ne 0$ inside the horizon, which is the a global minimum of the area radius $b$ and corresponds to a black-to-white hole transition.
\item There are two independent Dirac observables, which can be physically interpreted as the masses of the separate asymptotic Schwarzschild regions, {namely of the black and of the white hole masses respectively}.
\item The horizon is slightly smaller than the classical result (for black hole masses large w.r.t. the polymerisation scales).

\item {For black hole masses comparable or smaller of the polymerisation scales, there are large quantum effects at the horizon scale}.

\item There is an upper bound to the curvature, as measured for
instance by the Kretschmann scalar, and this value is obtained at the transition surface and depends on the black holes mass (as well as the white hole mass), unlike in the prior polymerization scheme of \cite{AshtekarQuantumTransfigurarationof} or in other models of non-singular black hole
\cite{Hayward:2005gi}
\end{enumerate}
This model has the nice feature that shows how an LQG--inspired quantization can lead to a resolution of the singularity and its explicit replacement by a (non-necessarily symmetric) black-hole-to-white-hole transition, a scenario suggested in \cite{Rovelli:2014cta}.
A possible shortcoming of the model is that the scale at which quantum effects become large depends on the parameters, and a certain restriction on the masses is necessary if one wants to confine this scale to inside the horizon.\footnote{This situation arises in other proposals as well \cite{BodendorferMassandHorizon}. An attempt to improve this situation has been made in \cite{BodendorferMassandHorizon,Bodendorferbvtypevariables}.}
On the other hand, the model is based on several arbitrary choices:
\begin{itemize}
\item[$(i)$] A specific foliation of the spacetime has been fixed, given by the level sets of the $r$ coordinate in Schwarzschild coordinates, and it has been shown that different foliations can give different physical results \cite{BojowaldAno-goresult}.
\item[$(ii)$] A specific  $\bar{\mu}$-scheme, based on the polymerisation of the two $P_i$ variables, neither more nor less. 
\item[$(iii)$] The polymerisation function for both variables is a sine function, in analogy with LQC models.
\end{itemize}
These choices are common, and indeed motivated by, the literature (see e.g. \cite{ModestoSemiclassicalLoopQuantum,CorichiLoopquantizationof,AshtekarQuantumExtensionof,AshtekarQuantumTransfigurarationof,BoehmerLoopquantumdynamics}).
However, more general recent proposals exist, see e.g. \cite{KellyEffectiveloopquantumgravity}.
Testing the robustness of the results with different choices, and conversely identifying choices that reduce the shortcomings, is crucial to move forward in the study of such models.
In the rest of this paper we focus on $(iii)$, and study the effect of changing the polymerisation function.

%-------------------------------------------------------
\section{Generalised polymerisation}
%-------------------------------------------------------

\subsection{Presentation of the setup}

To preserve the LQG-inspired idea of a polymer quantisation, it is sufficient to restrict the polymerisation function to a generic function (for a general discussion of this ambiguity see \cite{Perez:2005fn}, for its study in quantum cosmology see \cite{Amadei:2022zwp}).  Consequently, we consider a generic replacement
\begin{align}
P_i&\rightarrow \frac{f_i(\l_iP_i)}{\l_i},
\end{align}
where $f_1$ and $f_2$ are real, bounded, $C^1$ and periodic functions of periodicity $2\pi$, and such that 
$f_i(x) = x+o(x)$, that is
%$f_i(x)\overset{x\ll1}{\sim} x$. 
\begin{subequations}\label{eq:fcond}
	\begin{align}
		f_i(0) = 0, \label{eq:fcond1}
		\\
		f'_i(0) = 1, \label{eq:fcond2}
	\end{align}	
\end{subequations}
with the notation $f'_i=\frac{ \mathrm{d} f_i(\lambda_iP_i)}{\mathrm{d} \left(\lambda_iP_i\right)}$.
 Below in Section~\ref{asymptotic behavior} we will prove that these conditions are necessary in  order to recover the Schwarzschild solution in the large distance limit, but not sufficient: one further needs $f_2''(0)=0$. This additional condition however does not affect qualitatively the short distance structure, hence we will only introduce it when needed explicitly.  
Following this replacement, the new polymerized Hamiltonian is given by 
\begin{equation}\label{modified hamiltonian}
H_{\text{eff}}=\sqrt{n} \mathcal H_{eff}, \qquad \mathcal H_{\text{eff}} := 12v_1\frac{f_1(\lambda_1P_1)}{\lambda_1}\frac{f_2(\lambda_2P_2)}{\lambda_2}+4v_2\frac{f_2^2(\lambda_2P_2)}{\lambda_2^2}-\frac{1}{2}\approx 0,
\end{equation}
In the limit $\lambda_i P_i \rightarrow 0$ we recover the classical Hamiltonian, thanks to the required linear behaviour around the origin.

The polymerized dynamical equations are
\begin{subequations}\label{modified}
\begin{align}
\label{v1 modified}
\dot v_1&=12\sqrt{n}v_1f_1'\frac{f_2}{\lambda_2},\\
\dot v_2&=12\sqrt{n}v_1\frac{f_1}{\lambda_1}f_2'+8\sqrt{n}v_2\frac{f_2}{\lambda_2}f_2',
\label{v2 modified}
\\
\label{P1 modified}
\dot P_1&=-12\sqrt{n}\frac{f_1}{\lambda_1}\frac{f_2}{\lambda_2},\\
\label{P2 modified}
\dot P_2&=-4\sqrt{n}\frac{f_2^2}{\lambda_2^2},\\
\mathcal H_{\text{eff}} &=  3v_1\frac{f_1 }{\lambda_1}\frac{f_2 }{\lambda_2}+v_2\frac{f_2^2 }{\lambda_2^2}-\frac{1}{8}\approx 0.
\end{align}
\end{subequations}
They generalize \eqref{polymerised equations} in a straightforward way.
Remarkably, it is possible to solve these equations in full generality without specifying the functions $f_i$, even though in a partially implicit way. 
The knowledge of the implicit general solution will be sufficient to describe a large number of features of the modified spacetime generated by $f_1$ and $f_2$. 
The first step to construct this solution is to identify the Dirac observables.

%-------------------------------------------------------------
\subsection{Dirac observables}\label{Dirac observable}
%-------------------------------------------------------------

Dirac observables are gauge independent quantities, namely they Poisson-commute with all the constraints.
The only constraint of our model is $\mathcal H_{\text{eff}}$ (cfr. Eq.~\eqref{modified hamiltonian}), which also generates the dynamics.
Dirac observables will coincide with constants of motion, thus their identification leads immediately to the general solution of the equations of motions. More precisely, the expressions for the Dirac observables plus the constraints provide sufficient implicit relations to determine all phase space variables in terms of a chosen internal clock.

As the kinematical phase space is four-dimensional and we have one first class constraint, there are at most two independent Dirac observables. These were referred to as $K_i$ in the non-polymerized case, and we keep the same notation. They can be identified proceeding as in the non-polymerized case. The first can be found by inspection of the Hamiltonian constraint \eqref{modified hamiltonian} to be 
\begin{equation} \label{dirac observable1}
K_1:=v_1\frac{f_1}{\lambda_1}.
\end{equation}
Recall further, that $v_1$ is closely related to the area radius $b$ (cfr. Eq.~\eqref{cantransf}).
Therefore, this Dirac observable will be useful to analyse the area radius as a function of $P_1$ and to make statements about possible bounces and black-to-white hole transitions (see Sec. \ref{sec:bounces} below).
For the second Dirac observable again we divide \eqref{P1 modified} by \eqref{P2 modified} and integrate, obtaining as before

\begin{equation}\label{dirac observable2}
	K_2\left(P_1, P_2\right) = \lambda_1 \int_{1}^{P_1} \frac{ \mathrm{d}P_1}{f_1(\lambda_1P_1)}-3\lambda_2 \int_{1}^{P_2} \frac{\mathrm{d}P_2}{f_2(\lambda_2P_2)}.
\end{equation}
This constant of motion is well-defined for any choice of polymer function. However its explicit form can only be accessed once the functions are specified. The choice of lower bound in the integrals is made once and for all and it is needed to fix the freedom of constant shifts of Dirac observables. Here we picked 1 in agreement with the choice made in \eqref{K2}.

%------------------------------------------------------------------------------
\subsection{General expression of the metric}\label{General expression of the metric}
%------------------------------------------------------------------------------

To solve the dynamics in terms of the Dirac observables, we will start by noticing that the Hamiltonian constraint \eqref{modified hamiltonian} imposes that $f_2$ can never vanish on solutions. Consequently, the equation \eqref{P2 modified} implies that $P_2$ is a strictly decreasing function of $r$. It is thus a good internal clock, and we use it  to deparametrise the equations of motion,\footnote{Note that this is possible at the effective level as here we have access to $P_2$ directly. At the polymer quantum level $P_2$ is not available and this de-parametrisation would have to be reconsidered.}
and rewrite the metric coefficients as $a(P_2)$ and $b(P_2)$. Notice that with this choice the metric will be independent of the lapse, see \eqref{Karim}. The coordinate $r$ and $P_2$ are related via \eqref{P2 modified}
\begin{equation}\label{dr(dP2)}
dr=-\frac{\lambda_2^2}{4\sqrt{n}f_2^2}dP_2.
\end{equation}
Next, we want to determine the metric coefficients in \eqref{eq:canonicalmetric2} as functions of $P_2$.
To that end, $b$ can be expressed in terms of $P_1$ using the first Dirac observable \eqref{dirac observable1} and \eqref{v1(b)}, leading to
\begin{equation}\label{eq:bofP1}
b(P_1)=\left(\frac{3\lambda_1K_1}{2f_1}\right)^{1/3}.
\end{equation}
This can be turned into a function of $P_2$ using the implicit relation $P_1=P_1(\l_2P_2,K_2)$ determined by picking a specific value $K_2$ for the second Dirac observable \eqref{dirac observable2}. To make this step clear, we will use  the following notation,
\begin{equation}\label{eq:f1g1}
	f_1\big(\lambda_1P_1(\l_2 P_2,K_2)\big)=g_1(\lambda_2P_2,K_2)\ . 
\end{equation}
To determine $a$, we use the Hamiltonian constraint  \eqref{modified hamiltonian} to write
\begin{equation}
v_2=\frac{\lambda_2^2}{8f_2^2}-\frac{3K_1\lambda_2}{f_2}.
\end{equation}
Plugging this expression in \eqref{cantransf} we find
\begin{equation}\label{a(f1f2)}
a= \left(\frac{2 f_1}{3 \lambda_1 K_1}\right)^{2/3}\left(\frac{\lambda_2^2}{16f_2^2}-\frac{3K_1\lambda_2}{2f_2}\right).
\end{equation}
This way all metric components are expressed in terms of the Dirac observables $K_i$ and the internal clock $P_2$, and the metric \eqref{eq:canonicalmetric2} reads\footnote{Note that all factors of $n$ dropped out, which was expected as this is a pure gauge freedom specifying the coordinate $r$, which was replaced by the intrinsic clock $P_2$.}
\begin{align}
ds^2=&-\left(\frac{2g_1}{3K_1\lambda_1}\right)^{2/3}\left(\frac{\l_2}{4f_2}\right)^2
\left(1 - \frac{24 K_1f_2}{\l_2}\right)\dd\tau^2
\notag
\\
&+\left(\frac{3K_1\lambda_1}{2g_1}\right)^{2/3} \left(1 - \frac{24 K_1f_2}{\l_2}\right)^{-1} \l_2{}^2 \dd P_2{}^2+\left(\frac{3K_1\lambda_1}{2g_1}\right)^{2/3}\dd\Omega^2 \ , \label{modified line element}
\end{align}
with $\tau={t}/{\Lf}$ as before. 

Note that the Dirac observable $K_1$ appears explicitly in this line element, while $K_2$ enters only implicitly via the definition of $g_1$, see Eq.~\eqref{eq:f1g1}. Once the polymer functions $f_i$ have been specified, one can compute the integrals \eqref{dirac observable2} and obtain $P_1(P_2)$. This will make $g_1$ explicit and thus the metric.
It is quite remarkable that one can write the metric line element for completely arbitrary polymer functions, albeit with the limitation of an implicit function explained above. It means that one can deduce a great deal of the properties of the models without committing to specific choices too soon, and this is what we investigate next. As a sanity check, replacing the polymer functions by their arguments recovers the metric \eqref{Karim}, namely the Schwarzschild solution. 
The polymerized metrics on the other hand are obviously not solutions of the Einstein's equations, and as we show in the next Section, describe spacetimes with a varying number of horizons and bounces, depending on the polymer functions as well as the Dirac observables. A feature that remains from the unpolymerized model is that $K_1=0$ describes metrics everywhere  degenerate. In the following Section~\ref{asymptotic behavior}, we will show that even if the metrics do not describe a classical black hole, the Schwarzschild metric is recovered in the large radius limit, provided the condition $f_2''(0)=0$ is satisfied.

In order to do so, it is convenient to express the metric using the area radius as coordinate, as opposed to $P_2$. However, the polymerization makes $b(P_2)$ a non-monotonic function. This is known already for the sine functions \cite{BodendorferEffectivequantumextended}, and will become clear during the analysis of \eqref{modified} in the next Section. The area radius coordinate will have to be independently defined in different branches. On each of these branches, we have 
%\Sim{Fix this to be in the same form as the previous one} \Sam{Done}
\begin{equation}\label{line element with b}
		\dd s^2 =\, - \frac{1}{b^2}\left(\frac{\lambda_2}{4f_2}\right)^2\left(1 - \frac{24 K_1 f_2}{\lambda_2}\right) \dd \tau^2 +\frac{1}{f_1'^2}\left(1 - \frac{24 K_1 f_2}{\lambda_2}\right)^{-1}\dd b^2+ b^2 \dd \Omega^2
\end{equation}
In other words, \eqref{modified line element} is the maximal extension, and \eqref{line element with b} the expression valid in each area radius patch.

%----------------------------------------------------------------------
\section{Generic features of the dynamics}
%----------------------------------------------------------------------

In this Section we study the geometry of the polymer spacetime \eqref{modified line element}. We show that it is possible to work out several features such horizons, bounces, and singularity resolution without specifying the polymerisation functions. We start by analysing the equations of motion as a dynamical system, in order to identify the asymptotic regions and relative locations of the spacetime features.

%----------------------------------------------------------------------
\subsection{Evolution and fixed points}\label{evolution}
%----------------------------------------------------------------------
We proceed as follows. First, we eliminate the $v_i$ variables using respectively $K_1$ and the Hamiltonian constraint.
We then focus on the restricted space spanned by $P_i$, with equations of motion \eqref{P1 modified} and \eqref{P2 modified}.
Notice that $P_2\neq 0$, otherwise the Hamiltonian constraint is violated as a consequence of \eqref{eq:fcond}. We also exclude $P_1=0$, because again from \eqref{eq:fcond} it implies $K_1=0$ and as shown earlier these points describe degenerate metrics.

\begin{figure}[h!]
	\centering
	\subfigure[]
	{\includegraphics[scale=1.1]{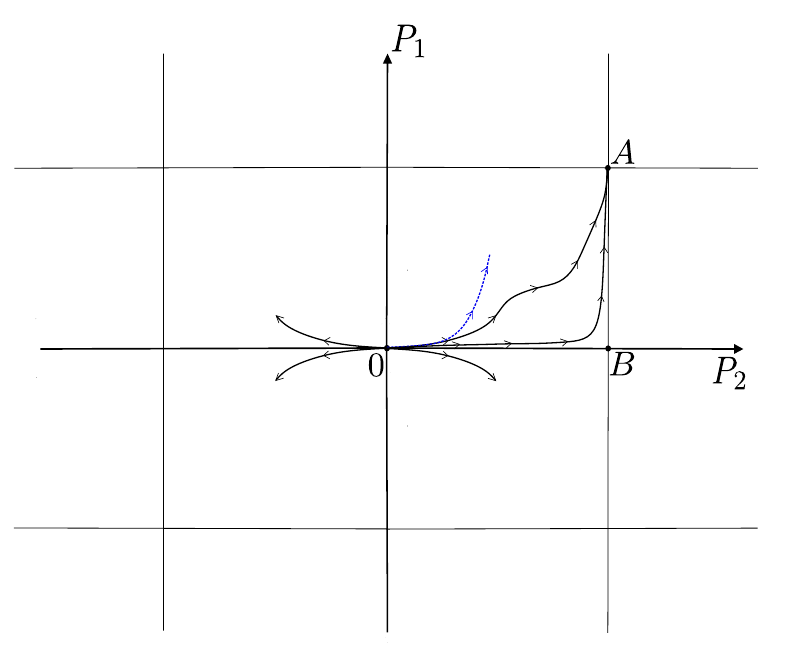}}
	\hspace{1mm}
	\subfigure[]
	{\includegraphics[width=7cm]{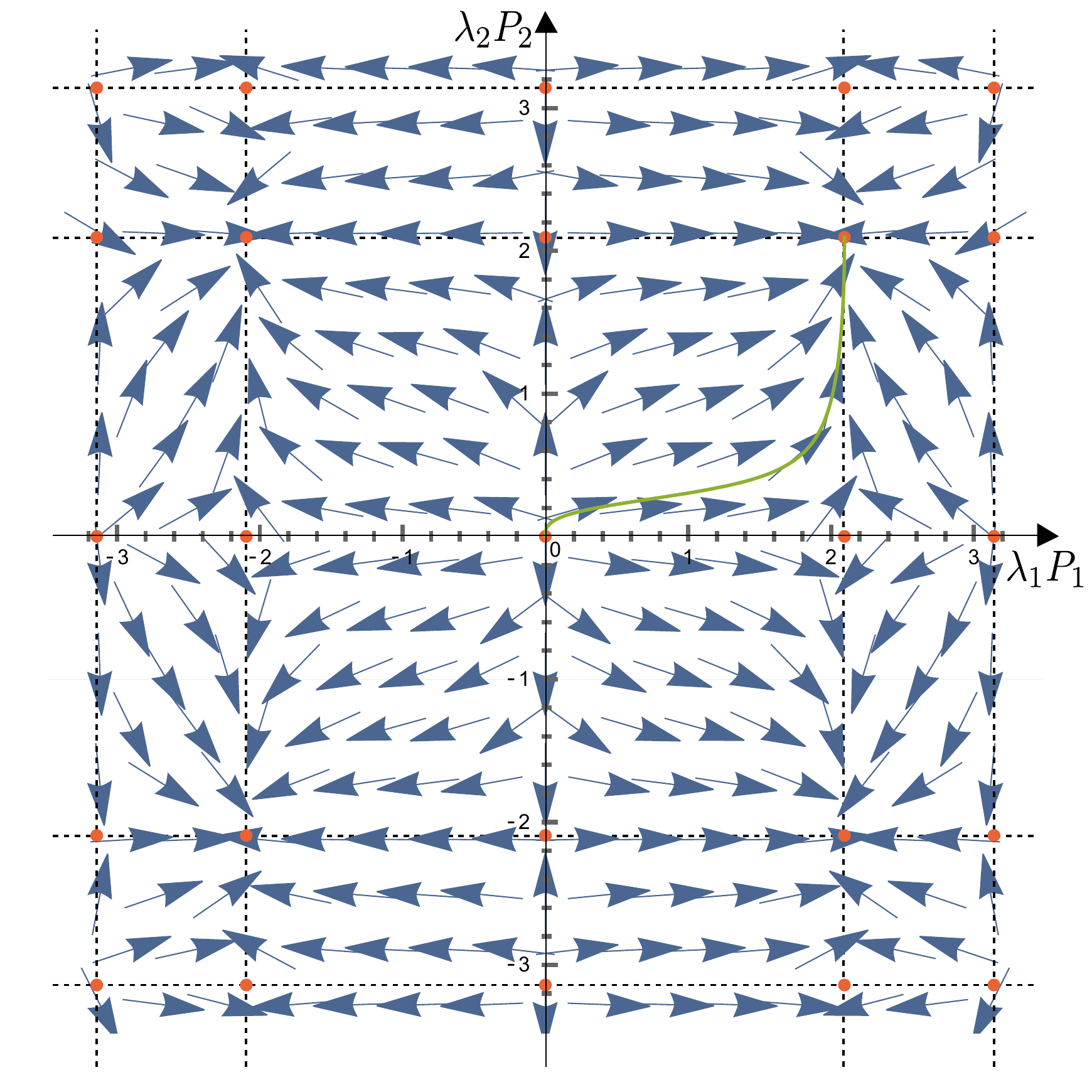}}
	\caption{(a) Qualitative picture of generic dynamics.  The horizontal and vertical straight lines are the zeros of $f_1$ and $f_2$ respectively. The trajectories are monotonic in both variables within each square. 
	The vertices of the check-board are alternatively repulsive and attractive asymptotic fixed points, like the origin and the point $A$, or unstable points like $B$.
	For reference, the blue dashed line is the relation in the case of the classical Schwarzschild solution.
	(b) Flow generated by the vector field $V$ 
with $f_1(x) = f_2(x) = \left(\sin(x) + \sin(2x)\right)/3$. In this example, an additional zero is present inside one period of each polymer function, hence the check-board of confined evolution has squares smaller than the period, and of different sizes. The (asymptotic) fixed points are marked with red dots. }

	\label{fig:flow} 
\end{figure}

To study the restricted space of the dynamics of the $P_i$'s, it is convenient to fix the lapse to 
\be\label{magicn}
\sqrt{n} = \frac{\lambda_2}{4 f_2},
\ee
because it decouples the equations of motion. This gauge choice is always accessible, as the Hamiltonian constraint~\eqref{modified hamiltonian} implies $f_2(\lambda_2 P_2) \neq 0$ throughout the evolution.
The dynamical flow is then described by the simple vector field 

\begin{equation}
	V(P_1, P_2) = \begin{pmatrix}
{\dot P_1} \\
{\dot P_2}
	\end{pmatrix} = %\stackrel{\eqref{modified}}{=}
	\begin{pmatrix}
		-3 {f_1}/{\lambda_1} \\
		-{f_2}/{\lambda_2}
	\end{pmatrix}\label{eq:flowfield}. % \quad , \quad \sqrt{n} = -\frac{\lambda_2}{4 f_2}
\end{equation}
We have already argued that $f_2$ can never vanish on solutions, because otherwise the Hamiltonian constraint is violated. On the other hand $f_1$ can vanish, but if it does, $K_1\equiv 0$, and the motion is then confined on points which describe only degenerate metrics. Excluding degenerate metrics, 
the polymer functions can never vanish along the solutions. 
This key property implies that there are no fixed points to the dynamical system, and that the $P_i$'s plane is partitioned into a check-board given by the zeros of the $f_i$'s, with the dynamics being trapped inside each square. See Fig.~\ref{fig:flow}, left panel, for an illustration. Notice that what matters is the zeros, not the period, of the $f_i$'s. If there are zeros before the period, one obtains an irregular check-board with some squares smaller than others, see the right panel of the figure. Furthermore, the evolution of the $P_i$'s is \emph{monotonic} between the extrema allowed. This allows us to get a pretty clear qualitative picture of the dynamics.
We focus on the four squares of the check-board connected to the origin, since they are the only ones that  contain the classical regime $\l_i P_i\ll 1$. Even though the origin in itself is not an allowed configuration, it plays the role of an asymptotic fixed point, as we now show. 

Near the origin the $f_i$'s have a linear behaviour, hence \eqref{eq:flowfield} gives $\sgn(P_i)=\sgn(\dot P_i)$. Not only the evolution is monotonic, but it is also tied to the sign of $P_i$. Therefore if we trace it backwards it will tend towards the origin, whichever of the 4 squares we are on. 
The origin must thus be a repulsive fixed point. It is actually an asymptotic fixed point, since the point itself is excluded from the phase space. The monotonic evolution without fixed points then forces the system to reach the farthest corner (across the diagonal of the square) for each square, see again Fig.~\ref{fig:flow}. These corners are thus attractive asymptotic fixed points.

The above description can be completed with an explicit perturbative expansion around the (asymptotic) fixed points. To that end, we posit
\be
P_i(r) = \os{P}_i+\epsy_i(r), 
\ee
where $\os{P}_i$ are zeros of $f_i$, and we Taylor-expand the polymer functions for $\l_i\epsy_i\ll 1$ around $\os{P}_i$. At leading order, \eqref{eq:flowfield}  simplifies to 
\be
\dot\epsy_1=-3\os{f}{}_1'\epsy_1, \qquad \dot\epsy_2=-\os{f}{}_2'\epsy_2,
\ee
whose solutions are
\be
\epsy_1=\os{\epsy}_1e^{-3\os{f}{}_1'\,r}, \qquad \epsy_2=\os{\epsy}_2e^{-\os{f}{}_2'\,r}.
\ee
From conditions \eqref{eq:fcond2} we conclude that the exponents are negative, meaning that the classical regime is reached for $r \rightarrow \infty$. Conversely, the farthest corner must have negative derivatives (because it is the next zero of the function), hence it is reached for $r\rightarrow -\infty.$ The remaining two corners of each square are unstable fixed points with one positive and one negative derivative.

\begin{figure}[h!t]
	\centering
	\includegraphics[scale=0.5]{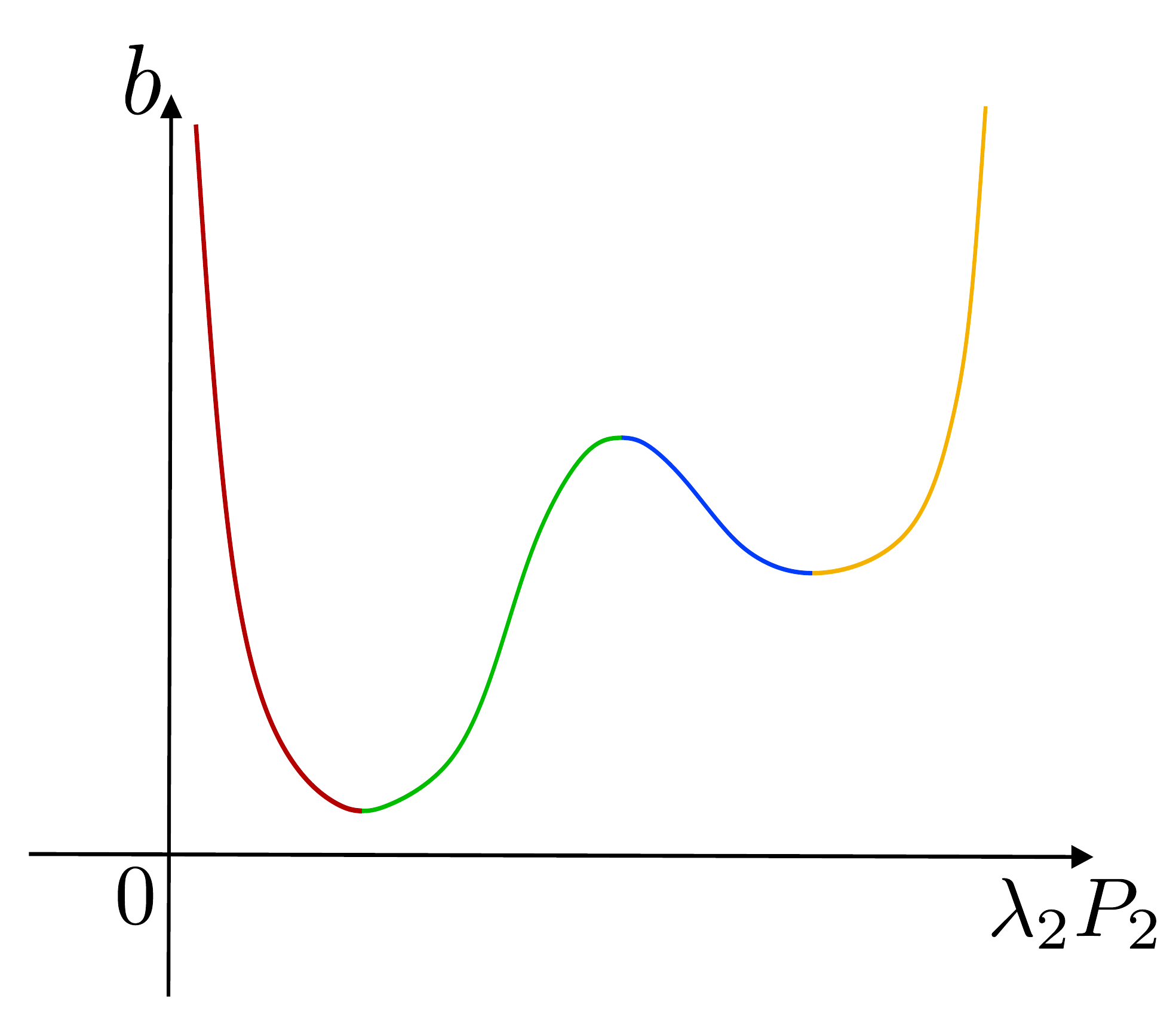}
	\caption{Plot of the area radius as a function of $\lambda_2P_2$. Different colours are used to separate the branches where the function $b(\lambda_2P_2)$ is invertible.}
	\label{b(l2P2)}
\end{figure}

This analysis can also be used to show that the area radius is not a monotonic function of $P_2$, and therefore not of $r$. In fact, we see from \eqref{eq:bofP1}, and the periodicity of $f_1$, that $b$  is not a monotonic function of $P_1$, and the monotonicity of the $P_i$ {as a function of $r$} implies that also $P_1(P_2)$ is monotonic.  Therefore, the inversion $P_2(b)$ can only be done for each of the branches, as illustrated in Fig.~\ref{b(l2P2)}. 

%---------------------------------------------------------------
\subsection{Bounces and singularity resolution}\label{sec:bounces}
%---------------------------------------------------------------

The notion of bounce can be visualized if we keep in mind the parallel between a black hole and a Kantowski-Sachs cosmological spacetime: a bounce is associated to a local minimum of the area radius $b(r)$. We will see in this Section that the location and number of bounces depends only on $f_1$. In the  next Section we will see that the location and number of Killing horizon is determined by $f_2$ instead.

The evolution equation for the area radius given by \eqref{v1 modified} is monotonic until we hit an extremum of $f_1$, namely a point such that 
\be
f_1'=0, \qquad f_1''\neq 0.
\ee
At this point the evolution turns around, and we have a bounce if it was initially decreasing, and a anti-bounce or turning point if it was increasing. 
Existence of at least one such point is guaranteed by the requirement of periodicity of $f_1$. So the bounce is guaranteed, but we can have as many bounces and counter-bounces (or turning points) as we want, choosing the appropriate $f_1$.
Furthermore, boundness of $f_1$ implies that that $b=0$ is only accessible for the solutions with vanishing $K_1$ and those are everywhere degenerate. Therefore the bounce must occur before reaching zero area radius. This suggests that the Schwarzschild singularity is resolved. To look into the question of singularities more precisely, one can evaluate the Kretschmann scalar associated with \eqref{modified line element}. This can be done explicitly with the aid of an algebraic manipulator like Mathematica or Maple. The result has the following form,
\begin{equation}\label{Kretschmann}
R_{\m\n\r\s}^2=\frac{F(g_1,f_2)}{f_2^4g_1^{8/3}},
\end{equation}
where $F$ is a six-order polynomial in $g_1$, $f_2$ and their first and second derivatives.
If we take  polymerisation functions that are periodic and $C^2$, the numerator in the previous expression is finite. The only divergences can arise from zeros of $g_1$ or $f_2$. These do occur, but at points that we have identified as the asymptotes for $r\rightarrow\pm\infty$.
Let us examine them separately.
The first of these is the origin in $P_i$ space (see Fig. \ref{fig:flow}) where we know that the solution describes the Schwarzschild metric at $r\rightarrow\infty$ with finite (indeed vanishing) Kretschmann invariant. For the opposite corners, we don't have a general argument for finiteness. However these points correspond to $r\rightarrow-\infty$, therefore in so far as this limit describes the post-bounce behaviour of another asymptotically flat spacetime, it will be non-singular.
We conclude that the only singularities allowed by these polymer models with smooth $f_i$'s are big-rip-type singularities, and require polymer functions such that the curvature diverges at $r\mapsto -\infty$.

Of course in the case where the spacetime is also a Schwarzschild spacetime around this fixed point, the Kretschmann scalar will be finite also near this fixed point. This case is particularly interesting since it corresponds to a black to white hole transition. We will see in the following how to choose the polymerization functions $f_1$ and $f_2$ to obtain this situation. This type of argument can in principle be extended to other curvature scalars and it is an indication of the absence of singularities away from asymptotic points in these models.

%----------------------------------------------------------
\subsection{Horizons}\label{horizons}
%----------------------------------------------------------

\begin{figure}[h!t]
	\centering
	\includegraphics[scale=0.5]{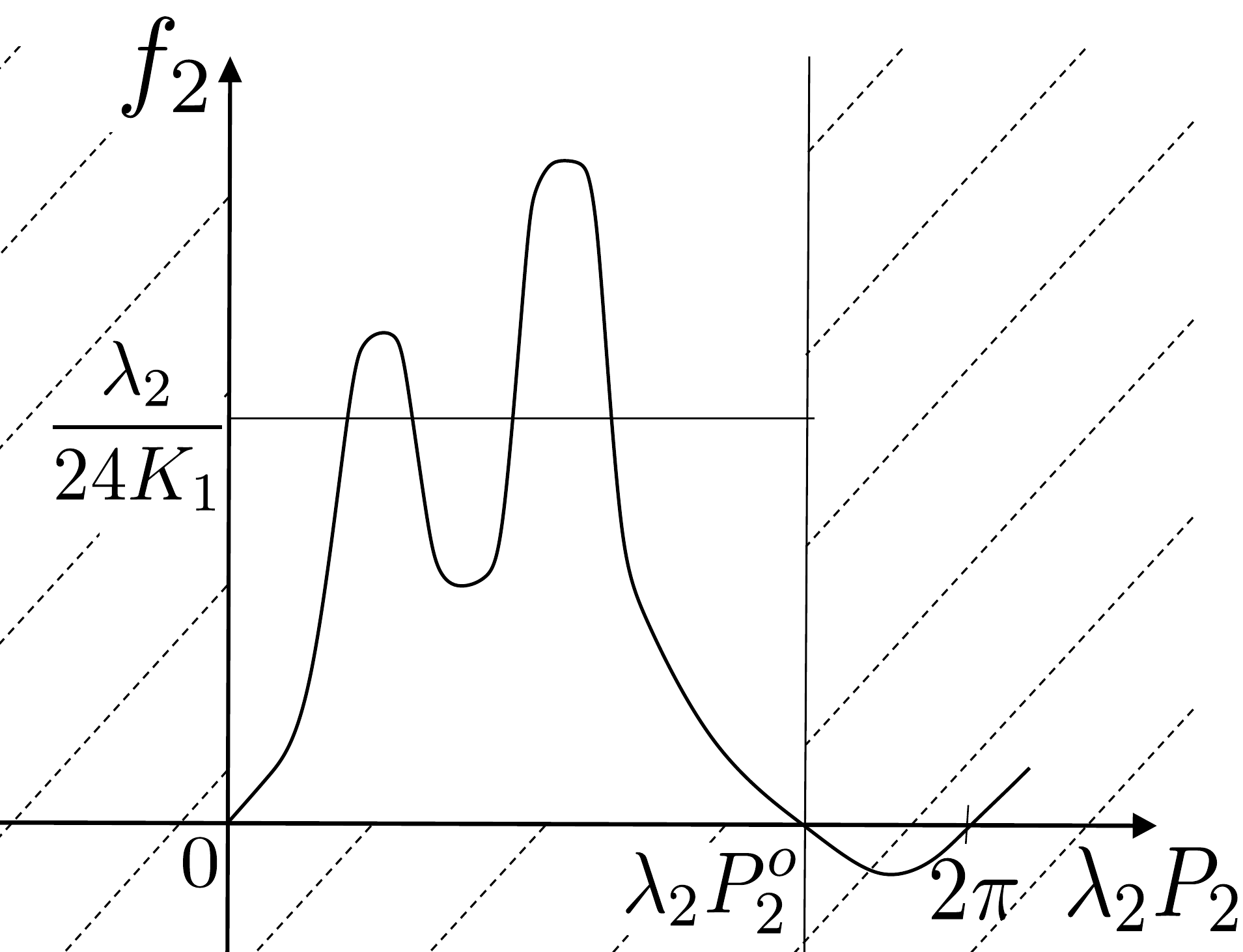}
	\caption{Sketch of an arbitrary example polymerisation function for $f_2$. Only the non-shaded region between $\lambda_2P_2=0$ and $\lambda_2P_2^o$ is relevant for the dynamics. The horizons correspond to the intersection of the function $f_2$ and the horizontal line $\frac{\lambda_2}{24K_1}$, explicitly depends on the chosen solution and the value $K_1$.}
	\label{fig:horcond}
\end{figure}

The number of horizons depends on $f_1$ alone, and their location on both $f_i$. From the general form \eqref{eq:canonicalmetric2} of the metric, we know that $\p_t$ is a Killing vector, therefore there is a Killing horizon whenever $a=0$. From \eqref{v2(a)} and the fact that $b \ne 0$ everywhere, this condition is equivalent to $v_2=0$.
Using then the Hamiltonian constraint \eqref{modified hamiltonian}, this translates to 
\begin{equation}
f_2=\frac{\lambda_1 \lambda_2}{24 v_1f_1} = \frac{\l_2}{24 K_1}.
\end{equation}
It is clear that it is possible to get any number of horizons,
depending on the choice of $f_2$ and the value of $K_1$. 
Fig.~\ref{fig:horcond} shows an example of polymer function admitting four horizons.
Since $f_2$ is bounded, there always exist values of $K_1$ for which there are no horizons. The bound in order to have at least one horizon is 
\begin{equation}\label{bound}
K_1\geq \frac{ \l_2}{24 \, {\rm max}f_2}.
\end{equation}
It follows that while the unpolymerized solution space describes always Schwarzschild black holes, of different mass and asymptotic time, the polymerized solution space always contains (non-singular) black holes as well as spacetimes without horizons (or exotic quantum stars).

A non-singular black hole scenario well considered in the literature is the black-to-white transition \cite{Haggard:2014rza,Rovelli:2014cta,Rignon-Bret:2021jch,BodendorferEffectivequantumextended,AshtekarQuantumExtensionof}, in which two event horizons sit on either side of a bounce. Sine polymer functions realize this scenario already \cite{BodendorferEffectivequantumextended}. One can then ask the question if the realization of this scenario imposes interesting restrictions on the polymer functions, and how much the scenario can be generalized allowing arbitrary polymer functions.
From the analysis just done, we see that the black-to-white transition corresponds to having a maximum of $f_1(P_1(P_2))$ between two intersections of $f_2(P_2)$ with $K_2$. No further requirement is needed. Therefore the freedom/ambiguity in the polymer functions cannot significantly be constrained by requiring that such scenario be realized. Conversely, all sorts of modifications of the scenario are possible by playing with global features of the regularization functions. The simplest qualitative novelty that can be introduced changing $f_i$ from the sine function is an asymmetry between the contracting phases. More in general, it is possible to combine multiple bounces and multiple horizons.

\begin{figure}[h!]
	\centering
	\subfigure[]
	{\includegraphics[scale=0.4]{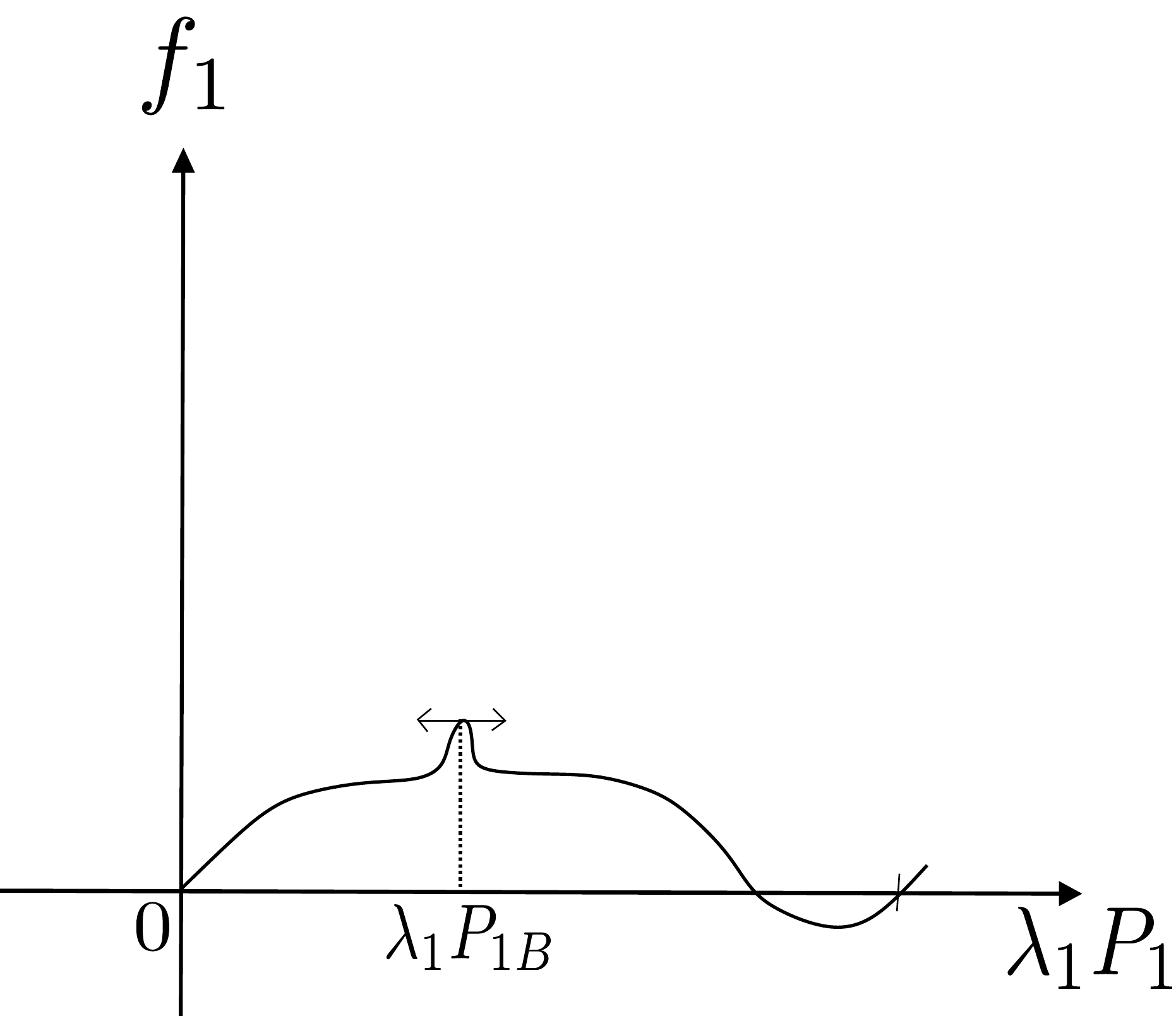}}
	\hspace{2mm}
	\subfigure[]
	{\includegraphics[scale=0.4]{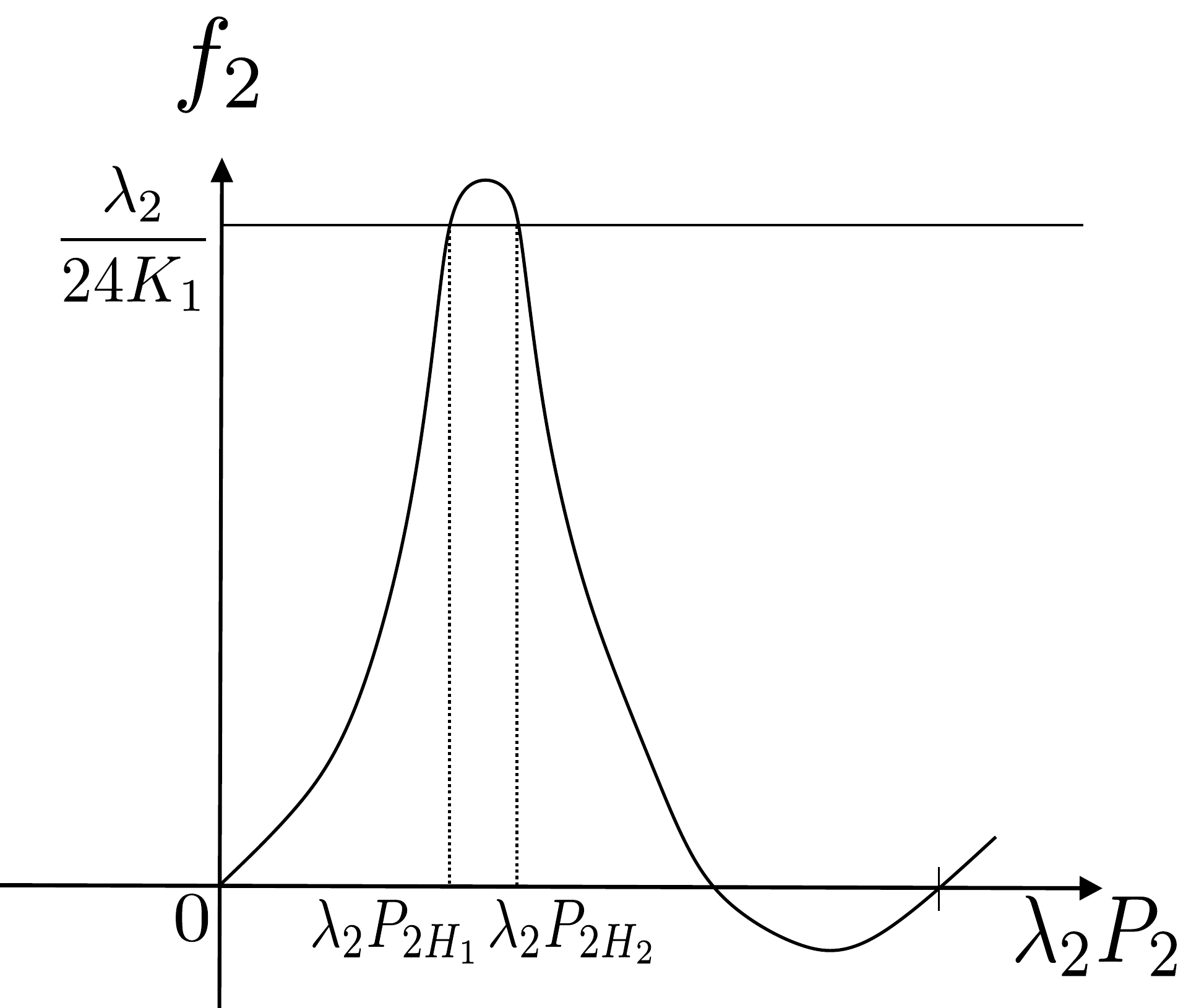}}
	\caption{(a) Plot of a specific polymerisation function for $f_1$ with only one local maximum between its two first zeros, i.e. only one bounce can appear (b) Plot of a specific polymerisation function for $f_2$ with only one local maximum between its two first zeros. Depending on the choice made for the Dirac observable $K_1$, at most two horizons will appear.}
	\label{fig:Figure3}
\end{figure}

\begin{figure}[t!]
	\centering
	\subfigure[]
	{\includegraphics[scale=0.4]{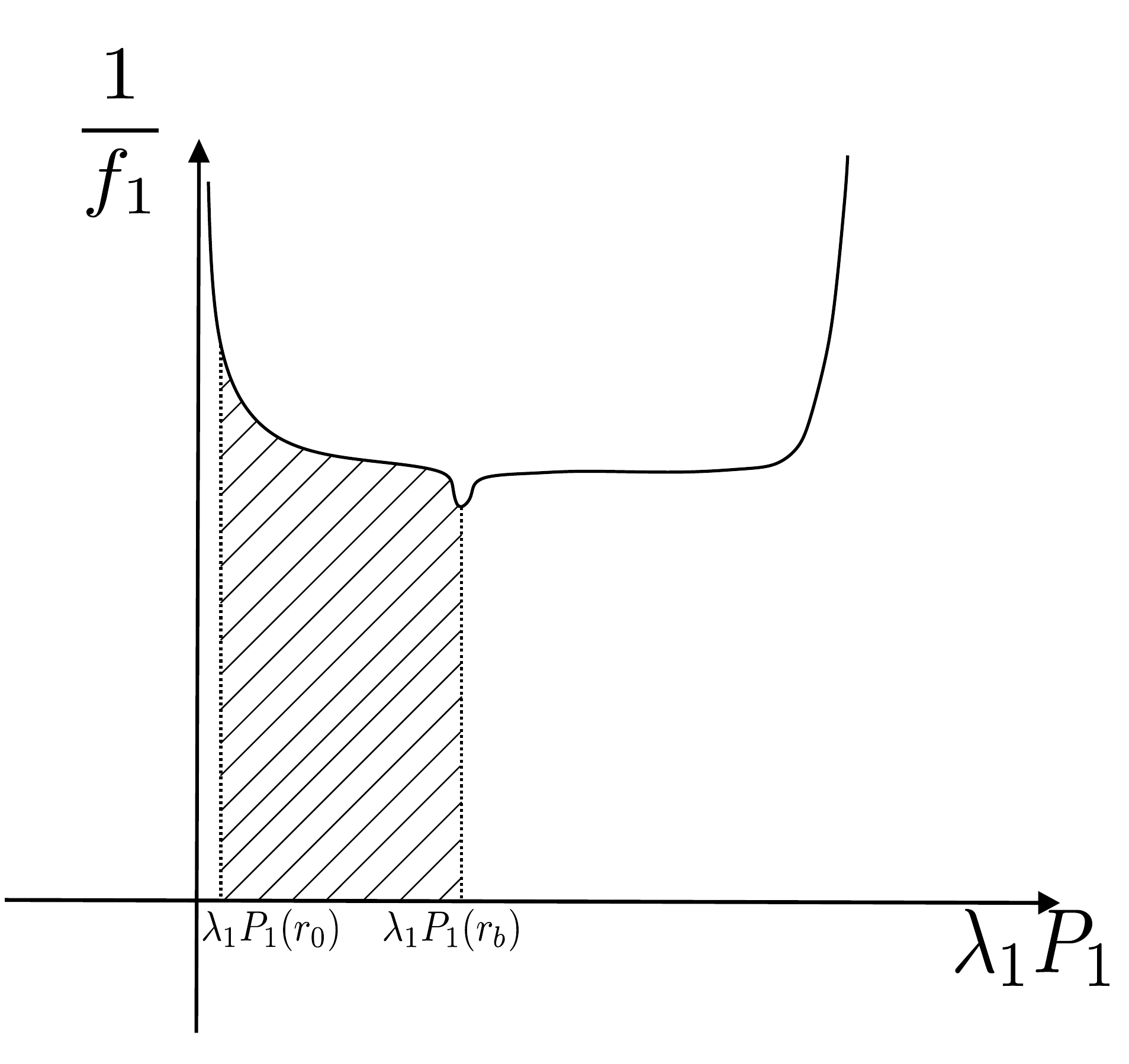}}
	\hspace{2mm}
	\subfigure[]
	{\includegraphics[scale=0.4]{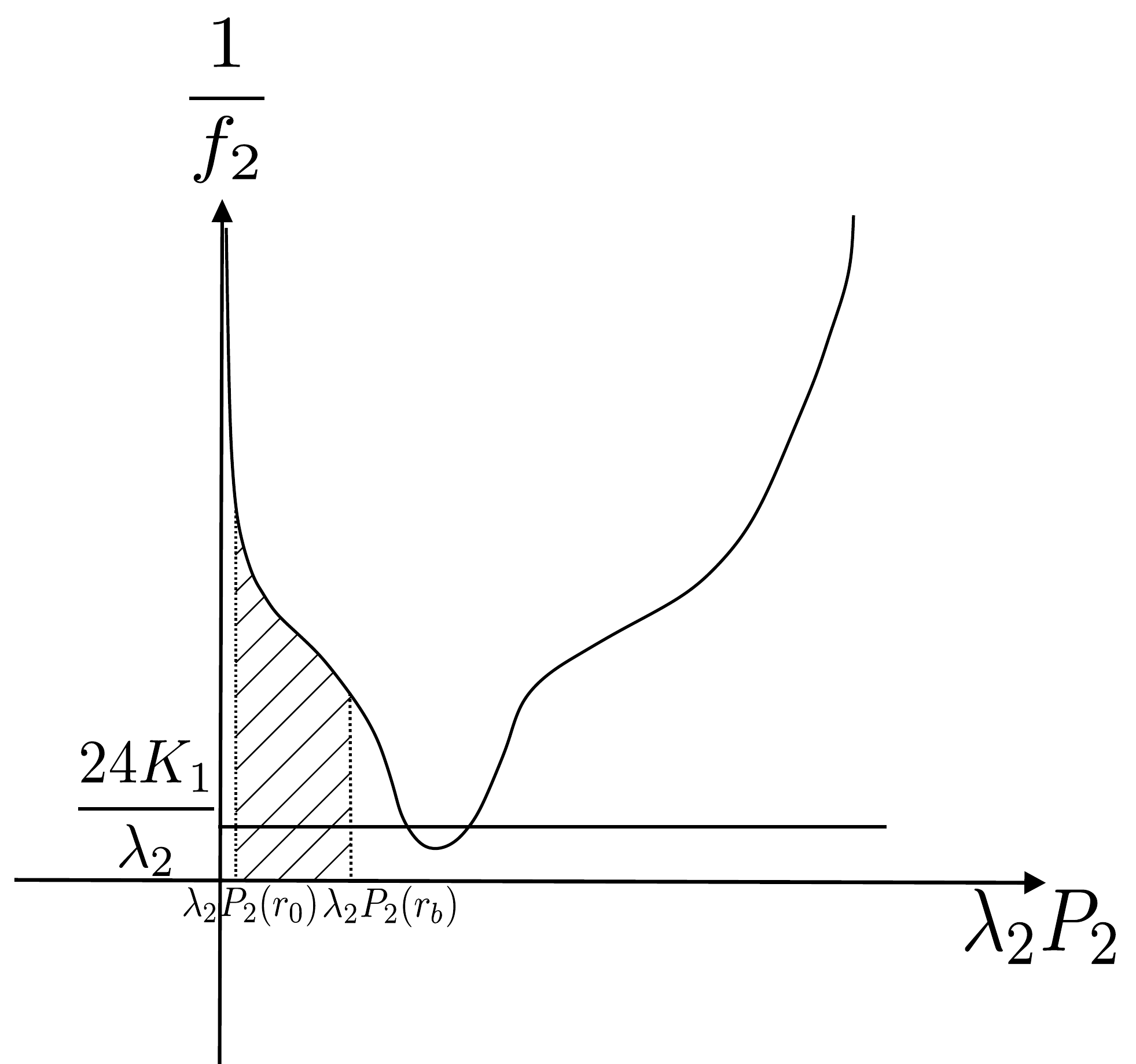}}
	\caption{Plot of the inverse of the polymerisation functions where the dashed area correspond to the integrals in \eqref{relation P1 and P2}.}
	\label{fig:Figure5}
	\label{fig:Figure6}
\end{figure}

In all cases, the relative location of bounces and horizons is not fixed, and will depend on both the choice of polymer functions and initial conditions. If the system is explicitly solved, the relative location can straightforwardly be determined from the functions $P_i(r)$. Interestingly, it is possible to find the relative location of bounces and horizons even without knowing the explicit solutions $P_i(r)$. This follows from the fact that the trajectories must satisfy the identity
\begin{equation} \label{relation P1 and P2}
3\int_{\lambda_2P_2(r_0)}^{\lambda_2P_2(r)}\frac{dx}{f_2(x)}=\int_{\lambda_1P_1(r_0)}^{\lambda_1P_1(r)}\frac{dx}{f_1(x)},
\end{equation}
as it follows from the constancy of the Dirac observable $K_2$ defined in \eqref{dirac observable2}. Given two reference values $P_i(r_0)$, this equation can be used to find the value of say $P_2(r)$ as a function of $P_1(r)$. 
To show how this is done in practise, consider the example of one bounce and two horizons, with polymer functions depicted in Fig.~\ref{fig:Figure3}. We want to establish if the bounce occurs in between the horizons, or outside. To bring out the dependence on the initial conditions, we can choose a value for $r_0$ in the classical domain, so that an approximate explicit solution is known, given by \eqref{P2 classical} and \eqref{v1 classical}.  Equation \eqref{relation P1 and P2}  fixes in this way  $P_2(r_b)$ as a function of $P_1(r_b)$ and classical inputs and can be used to determine the position of the bounce with respect to the horizons present in the effective geometry, as illustrated on Fig.~\ref{fig:Figure5}. All this depends explicitly  on the choice of the polymer functions and Dirac observables $K_1$ and $K_2$. The analysis holds for multiple bounces and multiple horizons.

%------------------------------------------------------------------------
\subsection{Trapped and anti-trapped regions}\label{causal structure}
%------------------------------------------------------------------------

In a black-to-white hole transition, the bounce separates a trapped and anti-trapped region between the two corresponding horizons.  We now show that the nature of the trapping or anti-trapping region turns out to be very simply determined by the sign of $f_1'$. More general cases, for instance if both horizons are on the same side of the bounce,  would require a systematic analysis.

\begin{figure}[t!]
	\centering
	\includegraphics[scale=0.5]{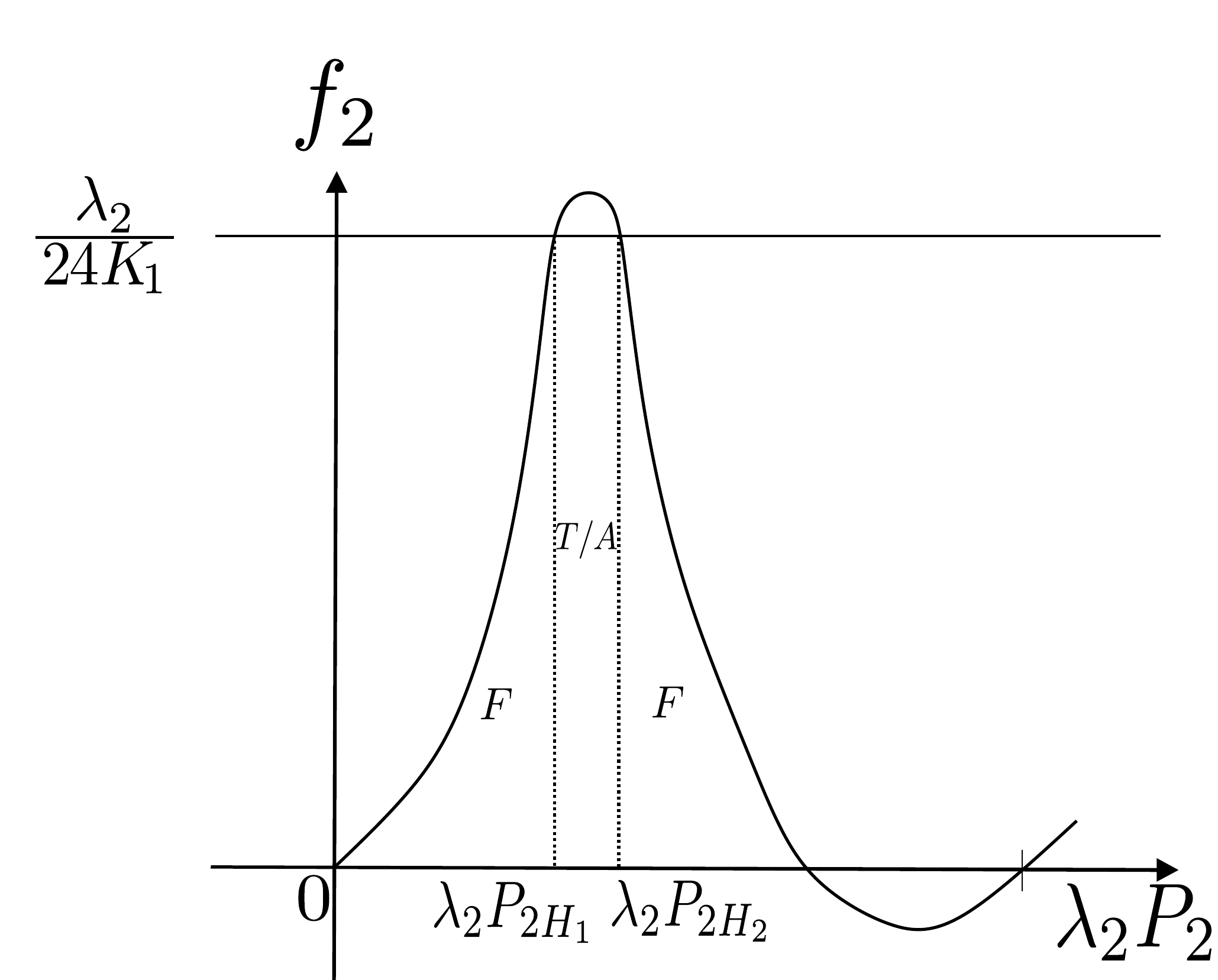}
	\caption{Plot of the polymerisation function $f_2$ chosen in Fig. \ref{fig:Figure3}. The horizontal line marks the threshold below which the region is free, and above which it is trapped/anti-trapped ($T/A$) depending on the sign of $\dot b$ as shown in Table \ref{fig:Table 1}. The intersections correspond to two horizons. If $f_1$ is chosen so to have a single bounce, three situations can occur. Starting from the asymptotic region at $r \rightarrow +\infty$, namely, $\lambda_2P_2=0$, the region between the horizon will be trapped (respectively anti-trapped) if the bounce occurs after the horizon (respectively before the horizon). If it occurs between the horizons, this region will be trapped and then anti-trapped.}
	\label{fig:Figure12}
\end{figure}

To see this, we recall that the outgoing and incoming expansions of null geodesic congruences for the spherical metric \eqref{line1} are given by (see e.g. \cite{AshtekarQuantumExtensionof,BodendorferEffectivequantumextended})
\begin{align}
\label{congruence +}
\theta_+&=\sqrt{-\frac{2}{N}}\frac{\dot b}{b}\text{sign}(a),\\
\label{congruence -}
\theta_-&=-\sqrt{-\frac{2}{N}}\frac{\dot b}{b}.
\end{align}
The sign of $a$ and $\dot{b}$ determine if the region is trapped, anti-trapped or free. We can summarize the possibilities in Table \ref{fig:Table 1}.
\begin{table}
\begin{center}
\begin{tabular}{|l|c|r|}
  \hline
    &$ \text{sign}(\dot b)=-$  & $\text{sign}(\dot b)=+$ \\
  \hline
  $\text{sign}(a)=-$ & Anti-trapped & Trapped \\
  \hline
  $\text{sign}(a)=+$ & Free & Free \\
  \hline
  	
\end{tabular}
\end{center}	\caption{Summary of the causal structure of a region in term of the sign of $a$ and $b$. These results come from \eqref{congruence +} and \eqref{congruence -}}
\label{fig:Table 1}
\end{table}
At this point, we now that regions outside the horizon ($a > 0$) are generically free regions.
Similarly, regions inside the horizon ($a < 0$) are trapped or anti-trapped, as illustrated in Fig.~\ref{fig:Figure12}.
The sign of $a$ is determined uniquely by the sign of $v_2$, see \eqref{v2(a)}, and the latter is positive for all $P_2$ such that
\be
f_2<\frac{\lambda_2 }{24K_1}.
\ee
However, which of both is the case, i.e. if the interior region is trapped or anti-trapped or if a transition happens as this depends on the position of the bounce relative to the horizons and thus the {polymer functions and the Dirac observables $K_1$ and $K_2$}.
To determine if the region between the horizon is trapped or anti-trapped, we have to look at $f_1$. 
We recall that since $\dot v_1 = 2\dot b b^2 = 12\sqrt{n}v_1f_1'\frac{f_2}{\lambda_2}$, and since $f_2>0$ we have $\text{sign}(\dot b )=\text{sign}(f_1')$.

\begin{figure}[h!]
	\centering
	\includegraphics[scale=0.5]{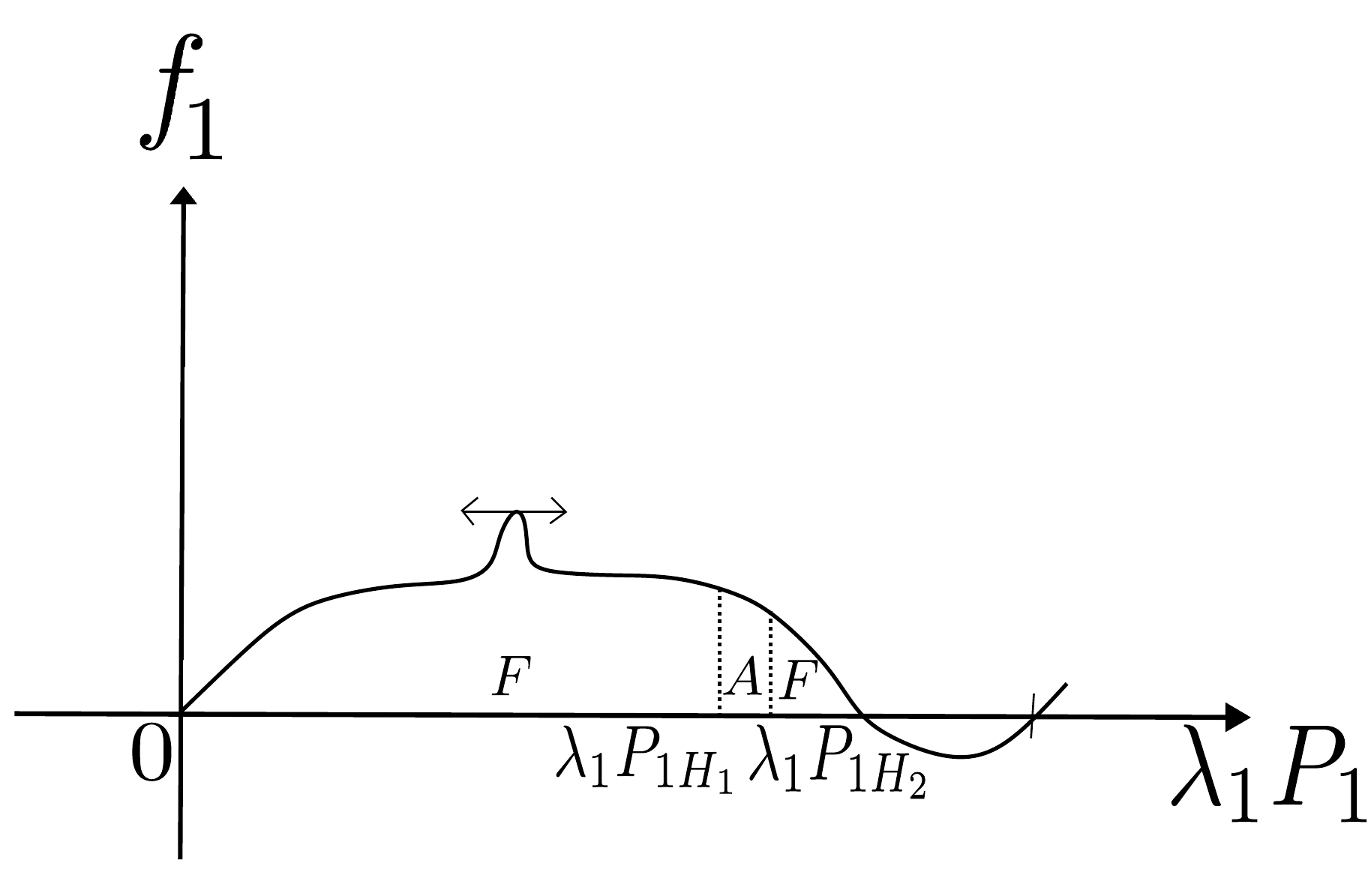}
	\caption{Plot of the polymerisation function $f_1$ chosen at the beginning of this section.The property free ($F$), trapped ($T$) or anti-trapped ($A$) of the spacetime is summarized on this plot.}
	\label{fig:Figure13}
\end{figure}

So in this specific case, since $f_1'<0$ between the horizons (cf Fig.~\ref{fig:Figure13}), which means that $\dot b<0$. So referring to Table \ref{fig:Table 1}, we deduce that the region between the horizons is anti-trapped. On the other hand, if we had chosen $C$ such that the bounce lies within the horizons, the bounce surface would have been a transition surface between a trapped and an anti-trapped region. In this case, a proper black-to-white hole transition appears.

%-------------------------------------------------------
\section{Large distance behaviour}
%-------------------------------------------------------

We have seen that the polymerized black holes present many non-classical features, such as multiple horizons and bounces. However, we would still like these solutions to reproduce the standard Schwarzschild black hole at large distances. This was the motivation to restrict the polymer functions to satisfy \eqref{eq:fcond}. These restriction were based on experience with loop quantum cosmology and with the standard polymer black hole. In this section we derive these conditions showing that they are indeed necessary in order to recover the Schwarzschild solutions at large scales. However, it turns out that they are not sufficient. One needs a further condition, given by $f_2''=0$ which is satisfied by the sine polymerization, for instance. 

Even thought this condition was not explicitly required in the analysis of the previous Section, a quick inspection shows that it was not used anywhere, hence all the results there presented apply to the class of polymer functions correctly reproducing Schwarzschild metric's at large distances. The analysis presented in this Section is also somewhat heavier and requires additional notation, which is our reason to leave it for the end.
It will also allows us to identify the mass of the black hole described by the generic solution \eqref{line element with b}.

An independent reason to study the large scale behaviour of the solutions concerns the interpretation of the spacetime % related question is to study the large scale limit 
\emph{after} the bounce. If asymptotic flatness is recovered in that limit as well, it then becomes meaningful to talk about the mass of the white hole, and discuss the physics of the bouncing process in terms of asymptotic charges.

Let us now come to the technical aspects. The area radius is given by the variable $b$, hence the large scale limit is $b\rightarrow \infty$.
However, while we can freely switch between $r$ and $P_2$ as time variables, $b(\l_2 P_2)$ is not a monotonic function, recall analysis of Section \ref{General expression of the metric} and Fig.~\ref{b(l2P2)}.\footnote{The origin of this is that $b(r)$ is no longer monotonic. This is why one can have $b$ inverting behaviour going through a bounce while $r$ runs smoothly over the whole real axis. Notice also that the value $r_{\sscr bounce}$ can occur for either positive or negative values, depending on the polymer function and the parameters of the solution.} Therefore, we have to restrict our analysis to each branch of invertibility. Since we are interested in the asymptotic behaviour, we can focus on the two branches connected to the $r\rightarrow\pm\infty$ limits which corresponds to $b\to \pm \infty$.
To study this, we will proceed with a perturbative expansion of the polymer functions with respect to $1/b$. Since the polymer functions depend implicitly on $b$ via the solution $P_i(b)$, this will be a nested perturbative expansion that requires the use of the equations of motion.

%------------------------------------------------------------------------------
\subsection{Asymptotic behaviour}\label{asymptotic behavior}
%------------------------------------------------------------------------------

Let us consider the regions in $P_i$-space connected to the origin $P_i=0$ (see Fig. \ref{fig:flow}). We denote $P_i^{\pm}$ the two asymptotic fixed points reached in the limit $r\to \pm \infty$ respectively (we assume that in the $r\to \infty$ limit we flow to the semiclassical regime, hence  $P^+_i=0$).
We introduce the following notation,
\begin{subequations}
\begin{align}\label{fixed points y}
&x:=\frac{1}{b} \ , \\
&z^\pm(x):=\l_1P_1%\left(\frac{1}{b}\right)
-\lambda_1P_1^{\pm}, \\
&y^\pm(x):=\l_2P_2-\lambda_2P_2^{\pm},
\end{align}
\end{subequations}
so to have $y^\pm$ and $z^\pm$ vanish at the two fixed points of each trajectory. We denote the fixed points $x=0^{\pm}$. 
To treat them at once with a single perturbative expansion we 
redefine the functions $f_i(\lambda_i P_i) \mapsto f_i(\lambda_i P_i - \lambda_i P_i^\pm)$, such that $f_i(0^\pm)=0$. 
In the following, we will drop the superscript $\pm$ from $y$ and $z$ to further lighten the notation.

In terms of these variables, the relevant components of \eqref{line element with b} read
\begin{subequations}\label{line element with x}
\begin{align}
	g_{\tau \tau}(x) =& - x^2\left(\frac{\lambda_2^2}{16 f_2\left(y(x)\right)^2} - \frac{3 K_1 \lambda_2}{2 f_2\left(y(x)\right)}\right)\;, 
	\\
	g_{xx}(x) =&\, g_{bb}(x)/x^4 \;.
\\
	g_{bb}(x) =& -\frac{\lambda_2^2 x^2}{16 f_1'(z(x))^2 f_2\left(y(x)\right)^2 g_{\tau \tau}(x)} \;,
\end{align}	
\end{subequations}
After some computations that can be found in Appendix \ref{Asymptotic expansion}, one can show that these metric components at first order in $x$ are 
\begin{subequations}\label{1st order metric}
	\begin{align} 
		g_{\tau \tau}&=-\frac{ \lambda_2^2}{16 f_2'(0^{\pm})^2y'(0^{\pm})^2}\left(1-\frac{2y'(0^{\pm})\left(12 K_1 f_2'(0^{\pm})^2 + \lambda_2 f_2''(0^{\pm})\right)}{\lambda_2 f_2'(0^{\pm})}x\right)+O(x^2) \ , \label{gttexp}\\
		1/g_{bb}&=f_1'(0^{\pm})^2\left(1-\frac{24 K_1 f_2'(0^{\pm})y'(0^{\pm})}{\lambda_2}x\right)+O(x^2) \label{gbbexp} \, .
	\end{align}
\end{subequations}
Since we have already fixed the area radius as coordinate, asymptotic flatness imposes
\be \label{2nd condi}
f_1'(0^{\pm})=\pm1.
\ee
Next, we still have the freedom fo rescaling $\t$, which we use to reabsorb the prefactor in $g_{\t\t}$, so to have asymptotic flatness in the new time variable without restrictions on the polymer functions. On the other hand, area radius also requires that the time and radial components are the inverse of one another. Inspection of the above equations shows that this occurs if and only if
\be \label{3rd condi}
\f{y'(0^{\pm})f''_2(0^{\pm})}{f'_2(0^{\pm})}=0 \qquad \Rightarrow\qquad f''_2(0^{\pm})=0
\ee
by the previous conditions.
In that case, the mass of the black hole is given by

\begin{equation}\label{Mass}
M = \frac{12 K_1 f_2'(0^{\pm})y'(0^{\pm})}{\lambda_2} \, .
\end{equation}
Notice that it depends on $K_2$ via $y'(0^{\pm})$, since $K_2$ appears in the relation between $P_2$ and $b$.\footnote{One may ask how this mass enters the Kretschmann scalar, given schematically by \eqref{Kretschmann}. The complexity of that expression makes it however hard to answer this question, even for the simplest choices of polymer functions.}

Consider now separately the two solutions for $r\to \pm \infty$. The corresponding values of $M$ will generically be different, because $f_2'(0^{\pm})$ and $y'(0^{\pm})$ can take different values.  Since $f_2'(0^{\pm})$ and $y'(0^{\pm})$ are both positive (respectively negative) in the black hole side (respectively in the white hole side), then the sign of the mass is only determined by the sign of $K_1$, as in the unpolymerized case.

In summary, in order to obtain the Schwarzschild solution asymptotically at the fixed point, we need to impose the conditions \eqref{1st condi}, \eqref{2nd condi} and \eqref{3rd condi}, i.e.

\begin{equation}\label{conditions}
	f_1'(0^{\pm}) = f_2'(0^{\pm}) \quad \text{and} \quad f_1'(0^{\pm})^2 = 1 \quad \text{and} \quad f_2''(0^{\pm}) = 0 \;.
\end{equation}
This fixes of the three of the four derivatives of $f_1$ and $f_2$ up to the second order expansion.
Note that in the case of the sine polymerisation, these conditions are automatically satisfied.

\begin{figure}[t!]
	\centering
	\includegraphics[scale=0.5]{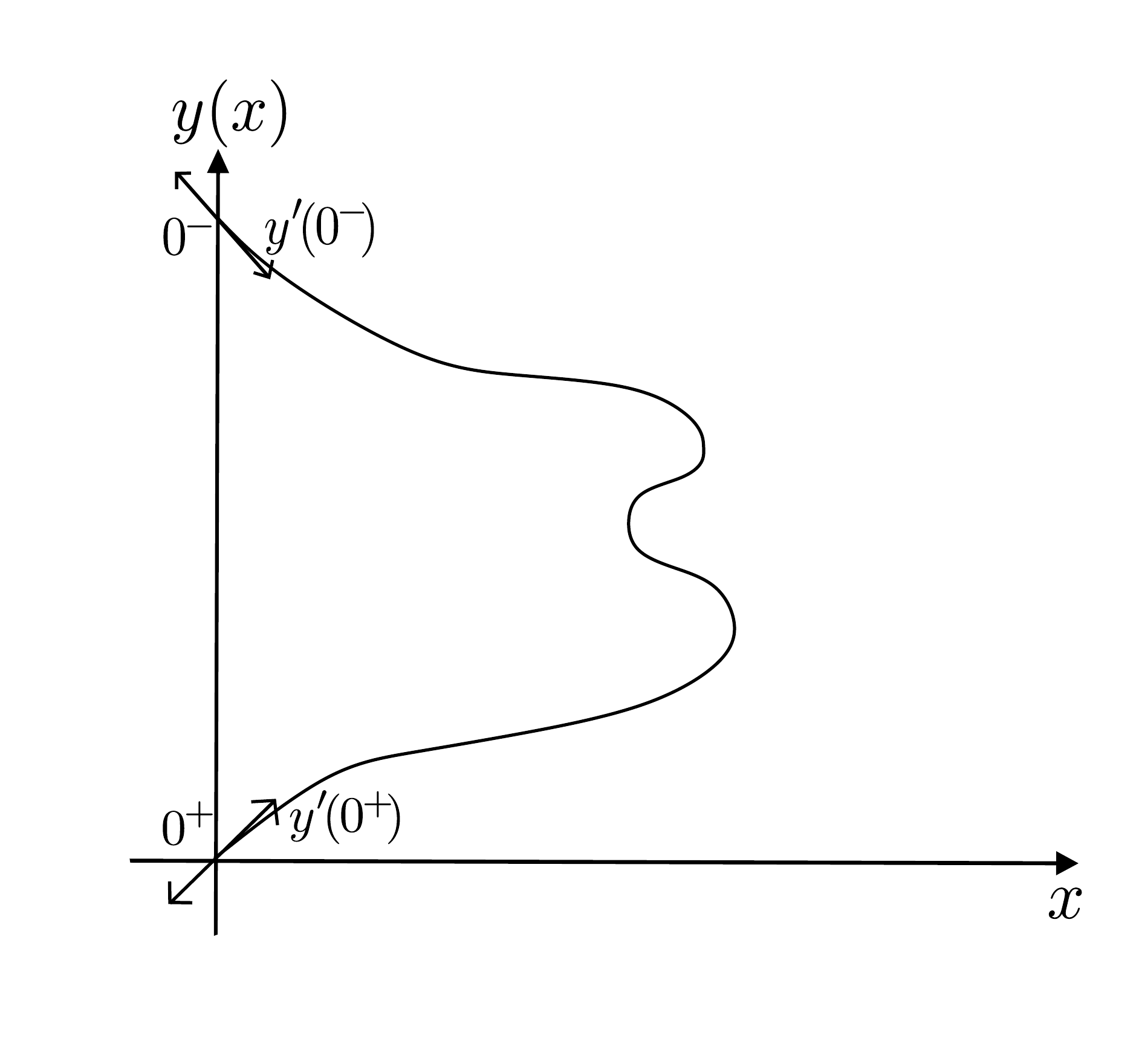}
	\caption{Sketch of the parametric solution $y(x)$. $y_-'(0^{\pm})$ uniquely fixes the trajectory $y(x)$ and thus the final derivative $y_+'(0^{\pm})$.}
	\label{fig:rely'}
\end{figure}

If we impose \eqref{conditions} at both fixed points, we obtain an evolution from an asymptotic Schwarzschild region to another asymptotic Schwarzschild region.
Let us call  the masses in the two regions $M_{BH}$ and $M_{WH}$.\footnote{These names have absolutely no meaning in terms of a black or white hole. Both asymptotic regions are Schwarzschild so both of these regions contain both, a black and a white hole. However, for the initial region, the black hole lies in the future and for the final region the white hole lies in the past. Depending on where the bounce is located w.r.t. the horizons (see the discussion in Sec.~\ref{causal structure}), an observer falling radially and freely starting in the initial region, will experience a black hole first, then see a transition from a trapped to an anti-trapped region, which the observer would call a white hole region until, she finds herself in an asymptotic exterior region of a Schwarzschild spacetime. This motivates the given notation, but there is no deeper physical meaning.}
They are given by
\begin{equation} \label{Mbhwh}
	M_{BH} = \frac{12 K_1 f_2'(\lambda_2 P_2^-)y'(0^{\pm})}{\lambda_2} \quad , \quad M_{WH} = \frac{12 K_1 f_2'(\lambda_2 P_2^+)y'(0^{\pm})}{\lambda_2} \, .
\end{equation}
From \eqref{conditions} we see that $\left|f_2'(\lambda_2 P_2^-)\right|=\left|f_2'(\lambda_2 P_2^+)\right|=1$. The evolution studied in Sec.~\ref{evolution} is such that no zeros of $f_2$ are crossed. Therefore, $f_2'(\lambda_2 P_2^-) = - f_2'(\lambda_2 P_2^+)$
and
\begin{equation}\label{massratio}
	\frac{M_{WH}}{M_{BH}} = -\frac{y'(0^{\pm})}{y'(0^{\pm})}.
\end{equation}
{Moreover, as it is illustrated in Figure \ref{fig:rely'}, $y'(0^{\pm})$ is positive in the black hole side and negative in the white hole side. This implies that the black hole mass and the white hole mass have the same sign}.
It is also possible to locally\footnote{The change of coordinates is given by \eqref{Mbhwh} (recall that $y'(0^{\pm})$ depends on $K_2$), and this may not be globally bijective. Consequently, we are limited to saying that this change of coordinates exists locally.} change coordinates in phase space from $K_i$ to $(M_{BH},M_{WH})$.

Because the two asymptotic regions are disconnected, different asymptotic masses does not mean a violation of energy conservation: it only means that the time-like Killing vector cannot be normalized to one at both asymptotes.
If one were interested in solutions with matching masses, one would be looking at a 1-dimensional subspace of the 2-dimensional phase space, determined by \eqref{massratio} once the polymer functions are chosen.

\subsection{Taming quantum gravity effects outside the horizon}

An obvious issue with non-singular black holes is to contain the deviations from general relativity not to spoil the macroscopic behaviour.
To that end, we look at the next order term in the expansion \eqref{gttexp}, and impose bounds on it.
This requires $y^{(3)}(0^\pm)$, which can be computed as before
from the equation of motion \eqref{ddyofx} and taking the limit $x\rightarrow 0$.
This leads to an recursive relation with $y^{(3)}(0^{\pm})$ on both sides, which can be solved to obtain
\begin{equation}
	y^{(3)}(0^{\pm}) = \frac{y'(0^{\pm})^3 f_2^{(3)}(0^{\pm})}{2 f_2'(0^{\pm})},
\end{equation}
where we use explicitly the condition $f_2''(0^{\pm})=0$.
We then find
\begin{subequations}\label{gexpO2}
\begin{align}
g_{\tau \tau}=-\frac{ \lambda_2^2}{16 f_2'(0^{\pm})^2y'(0^{\pm})^2}\left(1-2Mx - \text{sign}\left(f_2'(0^{\pm})\right) \frac{y'(0^{\pm})^2 f_2^{(3)}(0^{\pm})}{2\lambda_2} x^2\right)+O\left(x^3\right).
\end{align}
\end{subequations}
The $O(x^2)$ term of $1/g_{bb}$ vanishes, as a consequence of  $z''(0^{\pm}) = 0$ (cfr. \eqref{zofx}) and using the condition $f_2''(0^{\pm}) = 0$.

To keep the quantum corrections small outside of the horizon, we need the third term to be much smaller than the second for $b=x^{-1}\geq 2M$, namely 
\be
\frac{y'(0^{\pm})^2 f_2^{(3)}(0^{\pm})}{8\l_2M^2} \ll 1. 
\ee
This means that for given polymer functions, the condition is always satisfied for large enough black holes. Moreover, the expansion \eqref{gexpO2} allows us to obtain a criterion on the value of the area radius for which the deviations from the Schwarzschild metric become order one, given by 
\begin{equation}
b \leq \sqrt{\left|\frac{y'(0)^2 f_2^{(3)}(0)}{2\lambda_2}\right|}.
\end{equation}
A more precise characterization of the quantum corrections can be obtained studying the violations of the energy conditions, like for example in \cite{DeLorenzo:2014pta}.

\section{Conclusions}\label{Sec:conclusion}

We have shown that it is possible to investigate generic features of effective models describing spherically symmetric 
black hole models inspired by loop quantum gravity taking into account the large freedom in choosing the polymerization functions.
Most of the relevant global features of the solutions, as the location of horizons and the existence of bounces, are directly related to global features of the polymerization functions and the values of some Dirac observables or constants of motion.
More precisely, the features of the polymerization function $f_1$ encode the physics of bounces, while $f_2$ and its relation to certain constants of motion determines the location of the Killing horizons. Consistency with the classical regime does not strongly restrict the infinite dimensional ambiguity in the polymerization procedure. However, it is worth pointing out that in addition to the usual requirements on $f_i(0^{\pm})$ and $f_i^\prime(0^{\pm})$ that are derived from the naive continuum limit, our asymptotic analysis shows that in addition one needs to have $f_2^{\prime\prime}(0^{\pm})=0$. The number of bounces and the number of horizons can be freely specified by tuning the polymer functions and suitably choosing constants of motion.  With respect to the list of properties of the BMM model (listed in Section \ref{BMMp}) we conclude that all of them hold for an arbitrary polymerization except the shrinking of horizon. The deformation of the horizon can be freely controlled playing with the choice of polymer function. We notice in passing that the model also allows for solutions with no horizons at all, where quantum effects produce violations of the classical Einstein's  equations that can be seen as quantum stars made of `matter' of an entirely quantum nature. 
The existence of such  a large ambiguity in the construction of these models should not be surprising on general grounds \cite{Amadei:2022zwp, Perez:2005fn}. On the upper side this freedom could prove useful in the investigation of off-shell hypersurface deformation algebra \cite{Bojowald:2016hgh, Blohmann:2022yqo, Alonso-BardajiAnomaly-freedeformations}. Our analysis can also be applied to specific choices of non-sine polymer functions. It would be interesting to see whether it could be extended to different $\bar\m$-schemes like the one proposed in \cite{Alesci:2019pbs}, and then be able to study analytically the behaviour of the geometry post-bounce \cite{Alesci:2020zfi}.

\subsection*{Acknowledgements}
The work of JM was made possible through the support of the ID\# 61466 grant from the John Templeton Foundation, as part of the \textit{The Quantum Information Structure of Spacetime} (QISS) Project (qiss.fr). The opinions expressed in this publication are those of the author(s) and do not necessarily reflect the views of the John Templeton Foundation.

\appendix

\section{Boundary term}\label{Boundary term}

In this Appendix we recall the derivation of the minisuperspace Lagrangian \eqref{defL}. We consider a single boundary given by a $r$=constant hypersurfaces outside the horizon.  The total action including the Gibbons-Hawking-York boundary term is 
\begin{equation}
S[C]=\frac{1}{16\pi}\int dr\int_{C} d^3x\sqrt{-g}R - \frac{1}{8\pi}\int_{C}d^3x \sqrt{|h|}K.
%= \int dr \, L.
\end{equation}
With the ansatz \eqref{line1}, we have
\begin{equation}
\sqrt{-g}R=\sin(\theta)\left(2\sqrt{\bar n}-\frac{4 \bar b\dot{\bar a} \dot{\bar b}}{\sqrt{\bar n}}-\frac{2\bar a \dot{\bar b}^2}{\sqrt{\bar n}}+\frac{\bar b^2\dot{\bar a} \dot{\bar n}}{2 \bar n^{3/2}}+\frac{2\bar a \bar b \dot{\bar b} \dot{\bar n}}{\bar n^{3/2}}-\frac{\bar b^2\ddot{\bar a}}{\sqrt{\bar n}}-\frac{4\bar a \bar b\ddot{\bar b}}{\sqrt{\bar n}}\right),
\end{equation}
hence the Einstein-Hilbert term gives
\begin{equation}
\frac{1}{16\pi}\int_C d^3x \sqrt{-g}R=\frac{\ell_0\sqrt{\bar n}}{2}\left(\frac{\dot {\bar a} \dot{ \bar b} \bar b}{\bar n}+\frac{\bar a \dot{\bar b}^2}{\bar n}+1\right)-\ell_0\frac{\dd}{\dd r}\left(\frac{\dot{\bar a}  \bar b^2}{4\sqrt{\bar n}}+\frac{\bar a \bar b \dot{\bar b}}{\sqrt{\bar n}}\right).
\end{equation}
Coming  to the boundary term, 
the unit-normal to the $r$ foliation is, outside the horizon, 
\be
n=\sqrt{\frac{\bar n(r)}{\bar a(r)}}dr, \qquad n^2=1.
\ee
To this we associate the induced metric and extrinsic curvature
\begin{equation}
q_{\m\n}=g_{\m\n}-n_an_b, \qquad K_{\m\n}=q_\m^\r \na_\r n_\n=\frac{1}{2}\pounds_n q_{\m\n}.
\end{equation}
With the ansatz \eqref{line1}, we find 
\begin{equation}
q=-\bar a \bar b^4\sin(\theta)^2,
\qquad
K=\frac{\dot{\bar a}}{2\sqrt{\bar n \bar a}}+\frac{2\sqrt{\bar a}\dot{\bar b}}{\sqrt{\bar n}\bar b}.
\end{equation}
Hence the boundary term gives
\begin{equation}
\frac{1}{8\pi}\int d^3x \sqrt{h}K = \ell_0\left(\frac{\dot{\bar a}  \bar b^2}{4\sqrt{\bar n}}+\frac{\bar a \bar b \dot{\bar b}}{\sqrt{\bar n}}\right).
\end{equation}
Adding up, we recover \eqref{defL}.

\section{Asymptotic expansion of the metric components} \label{Asymptotic expansion}

In this Appendix, we provide details on the asymptotic expansion of the metric components \eqref{line element with x} leading to \eqref{1st order metric}. First, one can note that the only functions whose expansions are needed are $f_2(y(x))$ and $f_1'(z(x)) = \frac{\dd f_1}{\dd z}(z(x))$, while all dependencies of $f_1(z(x))$ have been replaced by the use of the Dirac observable $K_1$ and Eq.~\eqref{eq:bofP1}.
This also fixed the function
\begin{equation}\label{zofx}
	z(x) = f_1^{-1}\left(3 \lambda_1 K_1 x^3/2\right) \,.
\end{equation}
With these notations, the formal Taylor expansions are given by\begin{subequations}\label{expansions}
	\begin{align}
		f_2(y(x)) &= f_2'(0^{\pm}) y'(0^{\pm})\, x +\frac{1}{2} \left(f_2''(0^{\pm}) y'(0^{\pm})^2+f_2'(0^{\pm}) y''(0^{\pm})\right)\, x^2+ O\left(x^3\right), \\
		f_1'(z(x))&= f_1'(0^{\pm}) +f_1''(0^{\pm}) z'(0^{\pm})\, x +\frac{1}{2} \left(f_1^{(3)}(0^{\pm}) z'(0^{\pm})^2+f_1''(0^{\pm}) z''(0^{\pm})\right)\, x^2 +O\left(x^3\right)\\
		&\stackrel{\eqref{zofx}}{=} f_1'(0^{\pm}) + O\left(x^3\right),
	\end{align}
\end{subequations}
with $f_i^{(n)}$ denoting the $n$-th derivative w.r.t. $y$ or $z$, respectively, while $y' = \dd y / \dd x$ and respectively, $y''$, $z'$ and $z''$.
It was assumed and will be in the following that $f_i'(0^{\pm}) \neq 0$, which would also violate the classical limit.
These expressions are formal however, as the derivatives $y'(0^{\pm})$, $y''(0^{\pm})$ are not specified.
The derivatives $z'$ and $z''$ follow from \eqref{zofx} and we see that linear and quadratic order vanish exactly.
From the equations of motion \eqref{modified} follow the expressions
\begin{subequations}\label{1st terms y z}
	\begin{align} 
		&y'(x) =\frac{f_2(y(x))}{xf_1'(z(x))} \ ,  \\
		&y''(x) =\frac{f_2'(y(x))f_2(y(x))}{x^2f_1'(z(x))^2}-\frac{f_2(y(x))}{x^2f_1'(z(x))}-\frac{9K_1\lambda_1f_1''(z(x))f_2(y(x))x}{2f_1'(z(x))^3} \ ,  \label{ddyofx}
	\end{align}
\end{subequations}
Extracting $y'(0^{\pm})$ requires then
\begin{equation}\label{dy0}
	y'(0^{\pm}) = \lim_{x\rightarrow 0} y'(x) = \frac{f_2'(0^{\pm})}{f_1'(0^{\pm})} y'(0^{\pm}). 
\end{equation}
This implies
\be \label{1st condi}
f_1'(0^{\pm}) = f_2'(0^{\pm}),
\ee
with $y'(0^{\pm})$ unconstrained.\footnote{Recall that the fixed points are not part of the phase space but only reached asymptotically. This eliminates the possibility of solving \eqref{dy0} taking  $y'(0^{\pm}) \equiv 0$.}
Similarly, for the second derivative, we get
\begin{align}
	y''(0^{\pm}) = \lim_{x\rightarrow 0} y''(x) = \frac{y'(0^{\pm})^2 f_2''(0^{\pm}) + f_2'(0^{\pm})y''(0^{\pm})}{2 f_1'(0^{\pm})} \,,
\end{align}
which can be solved to obtain
\begin{equation}
	y''(0^{\pm}) = \frac{y'(0^{\pm})^2f_2''(0^{\pm})}{f_2'(0^{\pm})} \;.
\end{equation}
Here we again used the constraint $f_1'(0^{\pm}) = f_2'(0^{\pm})$.
This makes the expansions \eqref{expansions} explicit and allows to compute the metric functions, leading to
\begin{subequations}
	\begin{align} 
		g_{\tau \tau}&=-\frac{ \lambda_2^2}{16 f_2'(0^{\pm})^2y'(0^{\pm})^2}\left(1-\frac{2y'(0^{\pm})\left(12 K_1 f_2'(0^{\pm})^2 + \lambda_2 f_2''(0^{\pm})\right)}{\lambda_2 f_2'(0^{\pm})}x\right)+O(x^2) \ , \label{gttexp}\\
		1/g_{bb}&=f_1'(0^{\pm})^2\left(1-\frac{24 K_1 f_2'(0^{\pm})y'(0^{\pm})}{\lambda_2}x\right)+O(x^2) \label{gbbexp} \, .
	\end{align}
\end{subequations}

\providecommand{\href}[2]{#2}\begingroup\raggedright\endgroup

%\bibliographystyle{JHEPs.bst}

%\bibliography{library}

\end{document}